\documentclass[fleqn,usenatbib]{mnras}

\usepackage{newtxtext,newtxmath}

\usepackage[T1]{fontenc}
\usepackage{ae,aecompl}

\usepackage{deluxetable}
\usepackage{graphicx}	
\usepackage{amsmath}	
\usepackage{amssymb}	

\title[Stellar population on the mass-size plane]{SDSS-IV MaNGA: Global stellar population and gradients for about 2000 early-type and spiral galaxies on the mass-size plane}

\author[Li et al.]{Hongyu~Li$^{1,2}$\thanks{E-mail: hyli@nao.cas.cn}, Shude~Mao$^{3,1,4}$,
Michele Cappellari$^{5}$, Junqiang Ge$^{1}$, R.~J.~Long$^{1,4}$, \and Ran~Li$^{1, 6}$, 
H.J. Mo$^{7,3}$, Cheng Li$^{3}$, Zheng~Zheng$^{1}$, Kevin~Bundy$^{8}$ Daniel Thomas$^{9}$,
\and Joel R. Brownstein$^{10}$, Alexandre Roman Lopes$^{11}$, David R. Law$^{12}$ and Niv Drory$^{13}$
\\
$^{1}$National Astronomical Observatories, Chinese Academy of Sciences, 20A Datun Road, Chaoyang District, Beijing 100012, China\\
$^{2}$University of Chinese Academy of Sciences, Beijing 100049, China\\
$^{3}$Physics Department and Tsinghua Centre for Astrophysics, Tsinghua University, Beijing 100084, China\\ 
$^{4}$Jodrell Bank Centre for Astrophysics, School of Physics and Astronomy, The University of Manchester, Oxford Road, Manchester M13 9PL, UK\\
$^{5}$Sub-Department of Astrophysics, Department of Physics, University of Oxford, Denys Wilkinson Building, Keble Road, Oxford, OX1 3RH, UK\\
$^{6}$Key laboratory for Computational Astrophysics, National Astronomical Observatories, Chinese Academy of Sciences, Beijing, 100012, China\\
$^{7}$Department of Astronomy, University of Massachusetts, Amherst MA 01003-9305, USA\\
$^{8}$UCO/Lick Observatory, University of California, Santa Cruz, 1156 High St. Santa Cruz, CA 95064, USA\\
$^{9}$Institute of Cosmology \& Gravitation, University of Portsmouth, Dennis Sciama Building, Portsmouth, PO1 3FX, UK\\
$^{10}$Department of Physics and Astronomy, University of Utah, 115 S. 1400 E., Salt Lake City, UT 84112, USA\\
$^{11}$Departamento de Fısica, Facultad de Ciencias, Universidad de La Serena, Cisternas 1200, La Serena, Chile\\
$^{12}$Space Telescope Science Institute, 3700 San Martin Drive, Baltimore, MD 21218, USA\\
$^{13}$McDonald Observatory, The University of Texas at Austin, 1 University Station, Austin, TX 78712, USA\\
}

\date{Accepted 2018 February 2. Received 2018 January 15; in original form 2017 October 30}

\pubyear{2017}

\begin{document}
\label{firstpage}
\pagerange{\pageref{firstpage}--\pageref{lastpage}}
\maketitle

\begin{abstract}
We perform full spectrum fitting stellar population analysis and Jeans Anisotropic modelling (JAM) of the
stellar kinematics for about 2000 early-type galaxies (ETGs) and spiral galaxies
from the MaNGA DR14 sample. Galaxies with different morphologies are found to be located on a remarkably
tight mass plane which is close to the prediction of the virial theorem, extending previous results for ETGs.
By examining an inclined projection (`the mass-size' plane), we find that 
spiral and early-type galaxies occupy different regions on the plane, and their stellar population properties
(i.e. age, metallicity and stellar mass-to-light ratio) vary systematically along roughly the direction of
velocity dispersion, which is a proxy for the bulge fraction. Galaxies with higher velocity dispersions 
have typically older ages, larger stellar mass-to-light ratios and are more metal rich, which indicates
that galaxies increase their bulge fractions as their stellar populations age and become enriched
chemically. The age and stellar mass-to-light ratio gradients for low-mass galaxies in our sample tend
to be positive ($\rm centre<outer$), while the gradients for most massive galaxies are negative. 
The metallicity gradients show a clear peak around velocity dispersion $\log_{10} \sigma_{\rm e}\approx 2.0$,
which corresponds to the critical mass $\sim 3\times 10^{10}M_{\odot}$ of the break in the mass-size relation.
Spiral galaxies with large mass and size have the steepest gradients, while the most massive ETGs, 
especially above the critical mass $M_{\rm crit}\ga 2\times 10^{11} M_{\odot}$, where slow rotator ETGs
start dominating, have much flatter gradients. This may be due to differences in their evolution histories,
e.g. mergers.
\end{abstract}




\section{Introduction}
Early-type galaxies (ETGs) have been found to follow several scaling relations, for example the fundamental plane 
\citep{Djorgovski1987,Dressler1987}, which describes the relationship between velocity dispersion $\sigma$,
effective (half light) radius $R_{\rm e}$ and luminosity $L$ (or surface brightness $\mu$). There are similar 
relationships for the stellar mass plane \citep{Hyde2009} and the mass plane \citep{Cappellari2006, Bolton2007},
in which the luminosity is replaced by stellar mass and total mass, respectively. These scaling relations are 
related to the viral theorem \citep{Faber1987}. The edge-on view of these planes are thin, especially for the mass
plane \citep{Auger2010,Cappellari2013b}.

For the face-on view, however, galaxies with different properties may be located in different regions.  
\citet{Graves2009} and \citet{Graves2010} studied the age, metallicity and mass-to-light ratio of the galaxies 
on the fundamental
plane using the SDSS \citep{York2000} single fibre spectrum of quiescent galaxies, and found there are systematic 
variations of the stellar populations across the fundamental plane. \citet{Springob2012} performed a 
similar investigation using data from the 6dF galaxy survey. 

With integral field unit (IFU) data, e.g. $\rm ATLAS^{3D}$ \citep{Cappellari2011}, CALIFA \citep{Sanchez2012},
MASSIVE \citep{Ma2014}, SAMI \citep{Bryant2015} and MaNGA \citep{Bundy2015}, one can estimate the dynamical mass
much more accurately and study the mass plane relationship (e.g. \citealt{Cappellari2013b}).
\citet{Cappellari2013a} and \citet{McDermid2015} studied the 
distribution of the mass-to-light ratio, angular momentum, stellar population and star formation history 
on the mass plane for the 260 early-type galaxies in the $\rm ATLAS^{3D}$ survey. They found the ages, metallicity,
elemental abundance and gas content of galaxies vary systematically on the mass-size plane
\citep[see Fig. 22 of][]{Cappellari2016}.

Population gradients contain information on galaxy evolution, e.g. accretion, merger 
\citep{Hopkins2009,Matteo2009} and radial migration \citep{Roediger2012,Zheng2015}.
There are many previous studies focused on the gradients of galaxies, e.g. correlation between age, metallicity 
gradients and galaxy properties such as stellar mass, colour, velocity dispersion 
\citep{Mehlert2003,Sanchez-Blazquez2007,Koleva2009,Spolaor2009,MacArthur2009,
Kuntschner2010,Tortora2010,Rawle2010,Barbera2012,Kuntschner2015} and environments
\citep{Sanchez-Blazquez2006b,Tortora2012,Roediger2011,Zheng2017,Goddard2017}.  

In this paper, we use the galaxies from the MaNGA DR14 \citep{DR14} sample, Jeans anisotropic model (JAM) 
\citep{Cappellari2008}, and full spectrum fitting technique (pPXF, \citealt{Cappellari2004}) to study the
distribution of the stellar population properties (i.e. age, metallicity,
stellar mass-to-light ratio and their gradient) 
on the mass-size plane (the projection along $\sigma$ direction of the mass plane) for galaxies with
different morphologies. 
The structure of the paper is as follows. In Section~\ref{sec:data-model}, we describe
the MaNGA data (Section~\ref{sec:sample}), dynamical modelling (Section~\ref{sec:dynamical_modelling}) 
and stellar population synthesis model (Section~\ref{sec:sps}). In Section~\ref{sec:results}, we show our
results concerning the mass plane relationship (Section~\ref{sec:mp}), the distribution of the global population
properties (Section~\ref{sec:population_on-plane}) and the distribution of the 
population gradients on the mass-size plane (Section~\ref{sec:grident_on-plane}). In Section~\ref{sec:conclusion},
we summarize our results and draw our conclusions. We make use of a flat $\rm \Lambda CDM$ cosmology 
with $\Omega_{\rm m}=0.315$ and ${\rm H_0 = 67.3} \, {\rm km} \, {\rm s}^{-1} \, \rm{Mpc}^{-1}$
\citep{Planck2013}.

\section{data and models}
\label{sec:data-model}

\subsection{MaNGA data and galaxy sample}
\label{sec:sample} 
The galaxies in this study are from the MaNGA Product Launch 5 (MPL5) catalogue (internal release, nearly
identical to SDSS-DR14, \citealt{DR14}), which includes 2778 galaxies of different morphologies. We base our
galaxy morphologies on the \textit{Galaxy Zoo~1} \citep{Lintott2008,Lintott2011} by first matching the MPL5
sample with Table~2 of \citet{Lintott2011}. For galaxies with uncertain flags or
not in the table, we classify them by their S{\'e}rsic index \citep{Sersic1963} from the 
NASA-Sloan Atlas\footnote{\url{http://www.nsatlas.org/data}} (NSA) catalogue which is based on SDSS imaging
\citep{Blanton2011}. We take galaxies with $n_{\rm S\acute{e}rsic}>2.5$ as early type galaxies (ETGs) and 
the remainder as spiral galaxies.
We then visually check all the galaxies to adjust any misclassified galaxies and to exclude merging galaxies.
Galaxies with low data quality (with fewer than 100 Voronoi bins with signal-to-noise, S/N, greater than 10)
are also excluded. In total, we have 2110 galaxies in our final sample, with 952 ETGs and 1158 spirals. 
In order to test the effect of morphology classification, we also examine our results using a subsample with
intermediate sersic index ($2<n_{\rm S\acute{e}rsic}<3$) excluded (932 spiral galaxies and 898 ETGs in this
subsample), and find that our conclusions remain unchanged.

IFU spectra are extracted using the MaNGA data reduction pipeline \citep{Law2016},
and kinematical data are extracted using the MaNGA data analysis pipeline (K. Westfall
et al. 2017, in preparation). The data analysis pipeline extracts the kinematic data from
the IFU spectra by fitting absorption lines using the pPXF software \citep{Cappellari2004,Cappellari2017} 
with a subset of the MILES \citep{Sanchez-Blazquez2006,Falcn-Barroso2011} stellar library, MILES-THIN.  
Before fitting, the spectra are Voronoi binned \citep{Cappellari2003} to S/N=10.
Readers are referred to the following papers for more details on
the MaNGA instrumentation \citep{Drory2015}, observing strategy \citep{Law2015}, 
spectrophotometric calibration \citep{Smee2013,Yan2016a}, and survey execution and initial
data quality \citep{Yan2016b}.

\subsection{Dynamical modelling}
\label{sec:dynamical_modelling}

\begin{figure*}
\includegraphics[width=\textwidth]{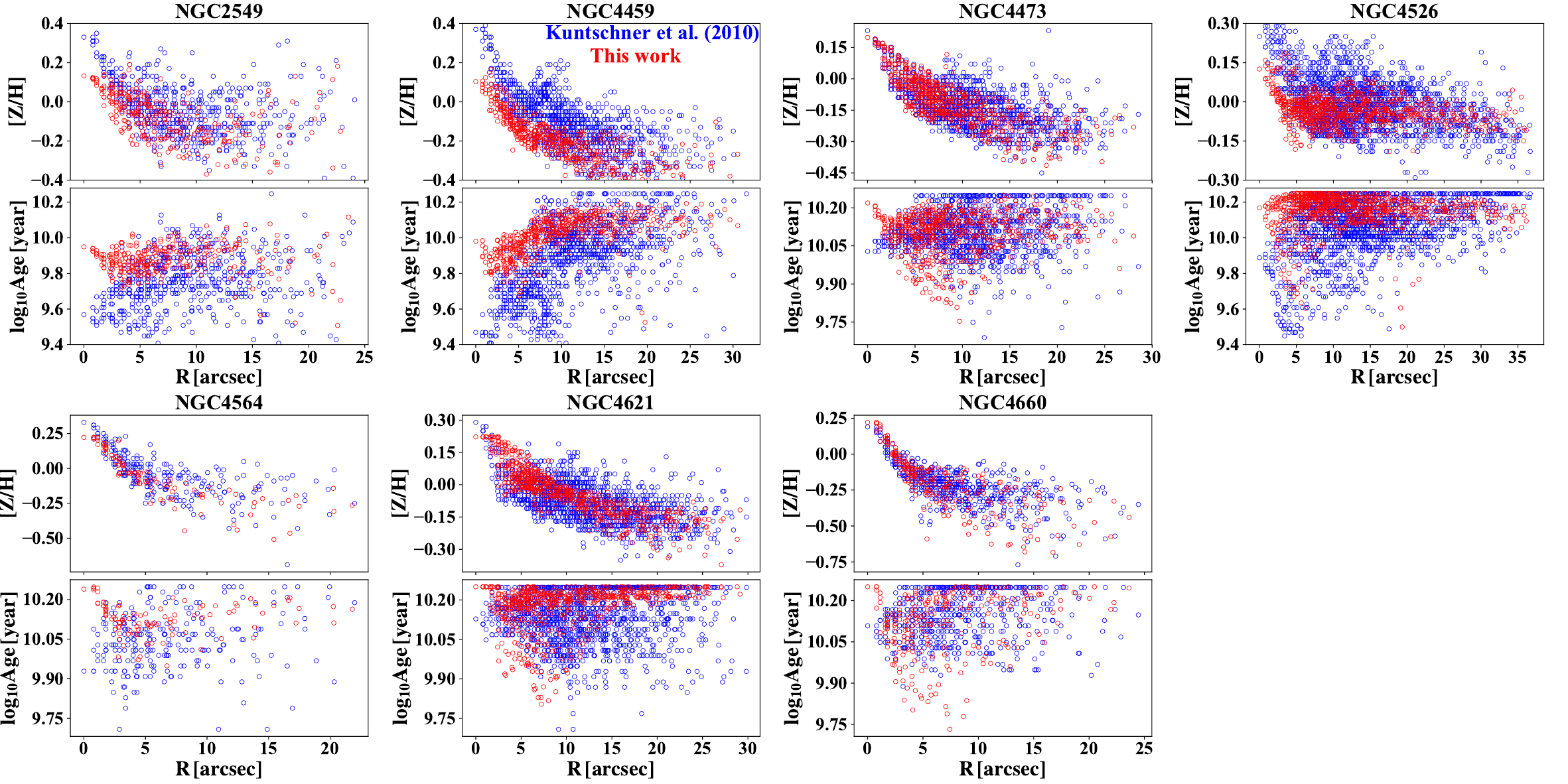}
\caption{Comparison of the logAge and [Z/H] profiles for 7 galaxies with high S/N from the $\rm ATLAS^{3D}$
         survey. The blue symbols are the results of each IFU bin from \citet{Kuntschner2010}, using line indices
         based method. The red symbols are the results of each IFU bin using the full spectrum fitting method
         described in Section~\ref{sec:sps}.
		}
\label{fig:a3d-cmp}
\end{figure*}

We perform Jeans Anisotropic Modelling (JAM, \citealt{Cappellari2008}) for all the 
galaxies in our sample. The modelling allows for anisotropy in the second velocity moments.
The total mass model has two components, i.e. a stellar mass distribution and a dark halo. For the 
stellar component, we first use the Multi-Gaussian Expansion (MGE) method \citep{Emsellem1994} with the 
fitting algorithm and Python software\footnote{Available from \url{http://purl.org/cappellari/software}}
by \citet{Cappellari2002} to fit the SDSS r-band image. We then deproject the 
surface brightness to obtain the luminosity density and assume a constant stellar mass-to-light
ratio to convert the light distribution to the stellar mass distribution.
For the dark matter halo, we assume a generalised NFW \citep{Navarro1996} halo profile
\begin{equation}
\label{eq:gnfw}
        \rho_{\rm DM}(r)=\rho_s \left(\frac{r}{R_s}\right)^\gamma
            \left(\frac{1}{2}+\frac{1}{2}\frac{r}{R_s}\right)^{-\gamma-3}.
\end{equation}
From running JAM within an MCMC framework ({{\bf emcee}, \citealt{Foreman2013}), we find the best-fitting
parameters which give the model best matching a galaxy's observed second velocity moment map.
The model gives a robust total mass estimation as demonstrated in \citet{Lablanche2012} and \citet{Li2016}
using numerical simulations.
Details of the modelling procedures can be found in \citet{Li2016,Li2017}. 

Following \citet{Cappellari2013b}, we calculate the size parameters $R_{\rm e}$, 
$R_{\rm e}^{\rm maj}$ and $r_{1/2}$ from the MGE models, and scale the $R_{\rm e}$ and
$R_{\rm e}^{\rm maj}$ by a factor of 1.35 \citep[see Fig.~7 of][]{Cappellari2013b}.
Here $R_{\rm e}$ is the circularized effective radius, $R_{\rm e}^{\rm maj}$ is the major axis
of the half-light isophote and $r_{1/2}$
is the 3-dimensional half-light radius. We define $M_{1/2}$ as the enclosed total mass within a spherical
radius $r_{1/2}$ from the best fitting JAM models. The velocity dispersion $\sigma_{\rm e}$ is defined as
the square-root of the luminosity-weighted average second moments of the velocity within an elliptical
aperture of area $A = \pi R_{\rm e}^2$
\begin{equation}
\label{eq:sigma}
\sigma_{\rm e} = \sqrt{\frac{\sum_{k} F_k(V_k^2+\sigma_k^2)}{\sum_{k} F_k}}
\end{equation}
where $V_k$ and $\sigma_k$ are the mean velocity and dispersion of the Gaussian which best fits the line-of-sight 
velocity distribution in the $k$-th IFU spaxel, and $F_k$ is the flux in the $k$-th IFU spaxel. The sum is within
the elliptical aperture described above. 
The $\sigma_{\rm e}$ so defined agrees quite closely with the velocity dispersion measured from a single
fit to the spectrum inside the same aperture \citep{Cappellari2013b}.

\subsection{Stellar population synthesis (SPS)}
\label{sec:sps}

\begin{figure}
\includegraphics[width=\columnwidth]{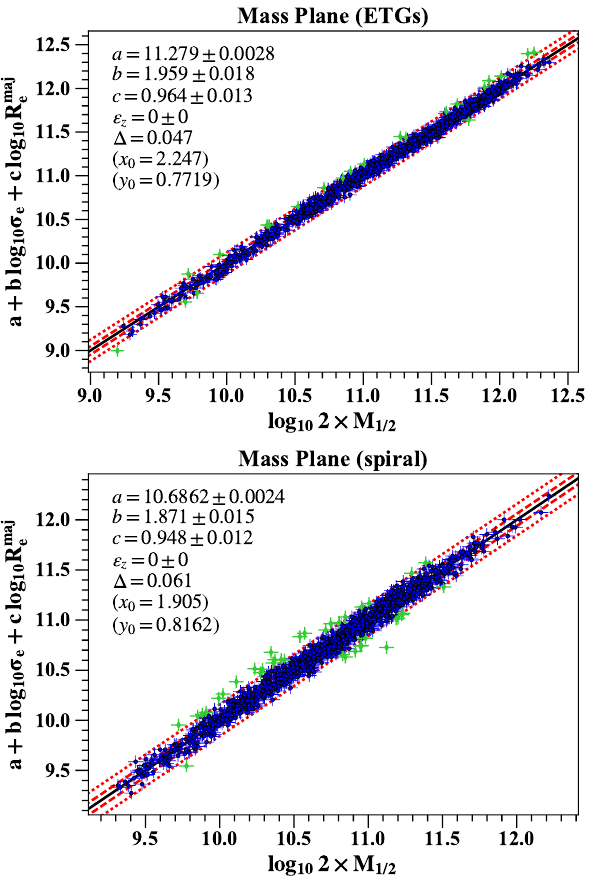}
\caption{The edge-on view of the best-fitting mass plane for the ETGs (top) and the spiral galaxies (bottom)
		 in our sample. The coefficients of the best-fitting plane $a+b(x-x_0)+c(y-y_0)$ and the observed scatter 
         $\Delta$ in $z$ are shown at upper left of each panel. The red dashed lines show the $1\sigma$ ($68\%$) 
         and $2.6\sigma$ ($99\%$). The outliers excluded from the fit by the {\bf LTS\_PLANEFIT} 
         \citep{Cappellari2013b} procedure are shown with green symbols.
		}
\label{fig:mp}
\end{figure}
We estimate the stellar population properties by fitting the MaNGA IFU spectra with stellar population templates.
Before spectrum fitting, we remove spectra with signal-to-noise ratio (S/N) less than 5 and bad sky 
subtractions. The data cubes are then Voronoi binned \citep{Cappellari2003} to S/N=30. We also used S/N=60,
and find that our results (age, metallity, stellar mass-to-light ratio and their gradients) are nearly unchanged.
The S/N for each spectrum is calculated as the ratio between the mean and the standard deviation of
the flux within a window from 4730\AA \ to 4780\AA, which does not include obvious emission and absorption lines.
We use the pPXF software \citep{Cappellari2004,Cappellari2017} with the MILES-based \citep{Sanchez-Blazquez2006}
SPS models of \citet{Vazdekis2010} and \citet{Salpeter1955} initial mass function.
We use 25 ages uniformly spaced in $\log_{10}$Age (years) between 7.8 and 10.25 and 6 metallicities 
([Z/H] = [-1.70, -1.30, -0.70, -0.40, 0.00, 0.22]). We assume a \citet{Calzetti2000} reddening curve
and do not allow for any polynomial and regularisation in the fitting. The fitting is performed between
$\sim 3500$\AA \ and $\sim 7400$\AA.
During the spectrum fitting we do not mask the gas emission lines in pPXF, but instead, we fit them 
simultaneously to the stellar templates as Gaussians, while adopting the same kinematics for all the gas emission
lines. We include the emissions from the Balmer series, the [OIII], [NII] doublet (with a fixed ratio $1/3$),
the [OI] doublet (with a fixed ratio $3/1$), the [OII] and the [SII].
We fit every spectrum twice. In the first pass, we fit all the good pixels and obtain the 
best fitting model spectrum. In the second pass, we remove all the pixels outside $3\sigma$ from the
first fitting. We use the results from the second fitting in our following analysis.

We calculate the luminosity weighted $\log_{10}$Age and metallicity [Z/H] using the equation below:
\begin{equation}
\label{eq:age-z}
\langle x \rangle = \frac{\sum_{j=1}^N w_j L_j x_j}{\sum_{j=1}^N w_j L_j},
\end{equation}
where $\rm w_j$ is the weight of the $j$th template and $\rm L_j$ is the corresponding r-band luminosity of the
SPS template. $x_j$ is the $\log_{10}$Age (or [Z/H]) of the $j$th template when calculating luminosity weighted
$\log_{10}$Age (or [Z/H]). The stellar mass-to-light ratio is calculated as 
\begin{equation}
\label{eq:ML}
M^*/L = \frac{\sum_{j=1}^N w_j M_j^{\rm nogas}}{\sum_{j=1}^N w_j L_j},
\end{equation}
where $M_j^{\rm nogas}$ is the stellar mass of the $j$th template,
which includes the mass in living stars and stellar remnants, but excludes the gas lost during stellar 
evolution. The other symbols are the same as in equation~\ref{eq:age-z}.

Having the $\log_{10}$Age, [Z/H] and $M^*/L$ for each bin, we take the luminosity weighted mean values within
one effective radius (an ellipse close to the half-light isophote) as the global population properties.
For the gradients of these properties, we first calculate the radial profiles of those quantities by taking
the median values in different elliptical annuli, with the global ellipticity measured around $1R_{\rm e}$.  
We then fit the profiles (logAge, [Z/H] and $\log_{10}M*/L$ vs.
$\log_{10} R/R_{\rm e}$) between $R_{\rm e}/8$ and $1R_{\rm e}$ to obtain the linear slopes as our gradients. 
A negative gradient means the central value is larger than the outer one.  

To test the robustness of our approach and to get a sense of the possible systematics, we apply our full
spectrum fitting method on 7 galaxies with high S/N from the 
$\rm ATLAS^{3D}$ survey, and compare our mettallicity and logAge profiles with the results from 
\citet{Kuntschner2010}, which are based on a radically different approach. The $\rm ATLAS^{3D}$ results are
in fact obtained by measuring three line indices, correcting them to a uniform velocity dispersion and then
locating the values, in a $\chi^2$ sense, on a 3-dimensional grid of individual SPS model predictions 
by \citet{Schiavon2007}, as a function of age, metallicity and alpha enhancement.
The comparisons are shown in Fig.~\ref{fig:a3d-cmp}. As one can see, two methods give comparable trends.
Our results show small scatters since we use full spectrum fitting rather than just line indices.
At the centre, our metallicities are slightly lower because our templates do not allow for non-solar abundances
and in particular do not have $\rm [Z/H]>0.22$ as in \citet{Kuntschner2010}. Correspondingly,
the central ages are slightly higher. Although the measurements were obtained with quite significant differences
both in the SPS model and in the fitting method, the agreement is quite good. In particular, the trends in both
the age and the metallicity gradients are consistent between the two approaches.

Throughout this work, we use the pPXF software combined with the MILES stellar template library
to obtain the stellar populations in galaxies. As a cross-check, we have also used the BC03 
\citep{Bruzual2003} SSP templates combined with the pPXF. With reasonable regularization, 
the BC03 library gives similar results as the MILES template. 

\section{Results}
\label{sec:results}

\subsection{Mass plane}
\label{sec:mp}
The mass plane relationship can be written as 
\begin{equation}
\label{eq:mp}
\log M_{1/2} = a + b \log \sigma_{\rm e} + c \log R_{\rm e}^{\rm maj},
\end{equation}
where $M_{1/2}$, $\sigma_{\rm e}$ and $R_{\rm e}^{\rm maj}$ are described in Section~\ref{sec:dynamical_modelling}.
The expected values from the virial theorem are $b=2$ and $c=1$ \citep{Faber1987}.
We fit this relationship separately for the ETGs and spiral galaxies in our sample, using the
{\bf LST\_PLANEFIT} procedure described in \citet{Cappellari2013b} which combines the Least Trimmed 
Squares robust technique of \citet{Rousseeuw2006} into a least-squares fitting algorithm which allows
for errors in all variables and intrinsic scatter. In the fitting, we assume $6\%$ error in
$\sigma_{\rm e}$, $6\%$ in $R_{\rm e}^{\rm maj}$ and $10\%$ error in $M_{1/2}$ \citep{Cappellari2013b}.
The results and the best fitting parameters are shown in Fig~\ref{fig:mp}. 

As one can see, both ETGs and spiral galaxies are on a remarkably tight mass plane, with coefficients $b$ 
and $c$ close to the prediction from the virial theorem. This agrees with the results of $b=1.942$, $c=0.991$ and
$\Delta=0.077$ in \citet{Cappellari2013a}, and follow the virial theorem slightly better than the results of
$b=1.67$, $c=1.04$, $\Delta=0.059$ from \citet{Scott2015} and $\alpha=1.857$, $\beta=-1.279$ in \citet{Auger2010},
with $\alpha=2$, $\beta=-1$ being the prediction from the virial theorem in their formalism. 
The observed scatter $\Delta$ for spiral galaxies is 0.061,
which is slightly larger than the value 0.047 for the
ETGs. This is because the uncertainties in measuring the velocity dispersion, effective radius and 
dynamical mass are larger for spiral galaxies due to the limited spectral resolution, the asymmetry of the galaxy 
and perturbation of the spiral arms. In the current data analysis pipeline for stellar kinematics, the extracted 
velocity dispersions under 50 km\,s$^{-1}$ have larger scatters, and may be slightly overestimated due to 
the uncertainties of the instrumental resolution. This may account for the slightly larger deviation of the
mass plane coefficients from the virial theorem for the spiral galaxies.
The intrinsic scatters $\varepsilon_{\rm z}$ for both ETGs and spiral
galaxies are consistent with being 0, until we reduce the error for the measured quantities in the fitting to 
$5\%$ ($\sigma_{\rm e}$), $5\%$ ($R_{\rm e}^{\rm maj}$) and $3\%$ ($M_{1/2}$). 
In the following sections, we use the `mass-size plane' to refer to the
projection of the mass plane along the $\sigma_{\rm e}$ direction. We choose this projection
because it is close to face-on and the two axes have clear physical meanings (i.e. mass and size).

\subsection{Stellar population on the mass-size plane}
\label{sec:population_on-plane}
We estimate the stellar population properties for all the galaxies in our sample using the full spectrum fitting method
described in Section~\ref{sec:sps}. The distributions of the velocity dispersions, ages, metallicities and 
stellar mass-to-light ratios of these galaxies on the mass-size plane are shown in Fig.~\ref{fig:sp-on-plane}. 
We use the Python implementation\footnote{Available from \url{http://purl.org/cappellari/software}}
\citep{Cappellari2013a} of the two-dimensional Locally Weighted Regression (LOESS, \citealt{Cleveland1988}) method to
obtain smoothed distributions, which are shown in Fig.\ref{fig:sp-on-plane_LOESS}.
The velocity dispersions on the mass-size plane agree well with the prediction from the virial theorem, as indicated
by the black dashed lines. The age, the metallicity and the stellar mass-to-light ratio change systematically on the 
mass-size plane for both ETGs and spiral galaxies. The values increase roughly along the velocity dispersion 
direction, which trace the bulge mass fraction \citep{Cappellari2016}. These systematic trends of the stellar population
with velocity dispersion are consistent with a picture in which the bulge growth makes the population more metal rich
and increases the likelihood for the star formation to be quenched \citep[see Fig.~23 of][]{Cappellari2016}. 

In a very recent work, \citet{Scott2017} showed similar results for $\sim 1300$
galaxies with different morphologies from the SAMI IFU survey. Other than our sample being slightly
larger (2110 vs. 1300), our two studies differ in four aspects
\begin{enumerate}
\item Their stellar population properties are derived from line indices, rather than full spectrum fitting 
      as in our study;
\item They use stellar mass while we use dynamical mass;
\item They study the galaxy global properties only while we study both global properties
      and gradients in the stellar populations (see Section~\ref{sec:grident_on-plane}).
\item Finally, the two studies are based on quite different samples, observed with different IFUs, and
      analyzed with different data pipelines.
\end{enumerate}

The 1-dimensional relationship between velocity dispersion and age, metallicity and stellar mass-to-light ratio
are shown in Fig.~\ref{fig:1d-sigma}. We fit the relationship using the equation below for ETGs and spiral
galaxies  
\begin{equation}
\label{eq:fit}
y=a+b(x-x_0),
\end{equation}
where $x_0$ is the median value of x. The best fitting line and coefficients are shown in each panel 
of Fig.~\ref{fig:1d-sigma}. The results from \citet{Scott2017} for their galaxies in clusters are shown in
Fig.~\ref{fig:1d-sigma} with black dashed lines. Their fitting did not separate ETGs and spiral
galaxies. Unlike Fig.~5 of \citet{Scott2017}, our logAge--velocity dispersion relation shows similar 
bimodality to the metallicity--velocity dispersion relation, and their [Z/H] reaches values as low as $-2$,
while in our results the values never go below $-1.4$. 

\begin{figure*}
\includegraphics[width=\textwidth]{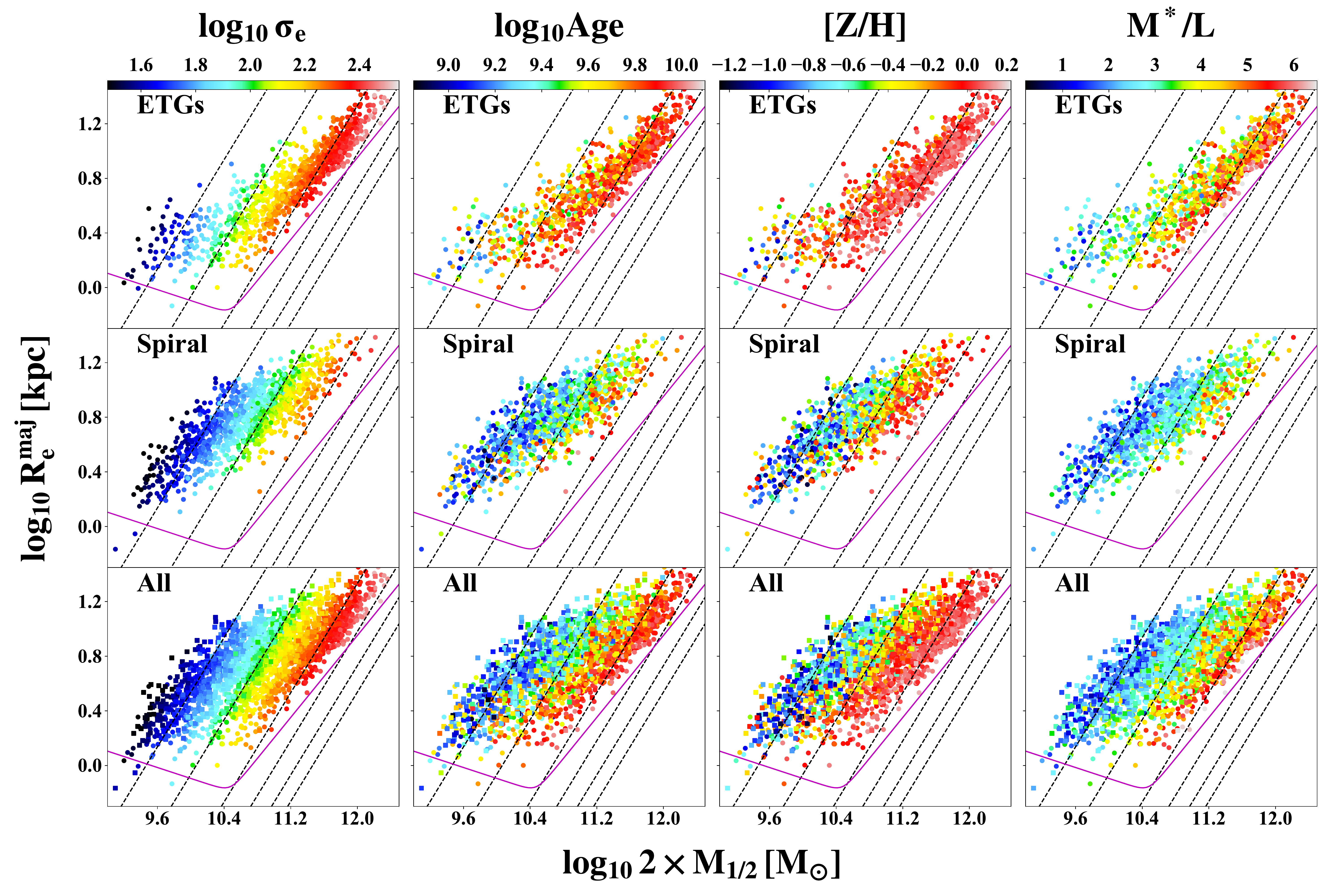}
\caption{Velocity dispersion $\sigma_{\rm e}$, logAge, metalicity [Z/H] and stellar mass-to-light ratio $M^*/L$ (SDSS
		 r-band) distribution on the mass-size plane ($R_{\rm e}^{\rm maj}$ vs. dynamical mass $M_{1/2}$). Colours 
         indicate the parameters as labelled at the top of each column. The results for early-type, spiral and all 
         galaxies are shown in the upper, middle and bottom panels, respectively. In the bottom panels, coloured squares 
         represent spiral galaxies and coloured circles represent ETGs. In each panel, dashed lines
         show lines of constant velocity dispersion: $50$, $100$, $200$, $300$, $400$, and $500$\,km\,s$^{-1}$ from
         left to right, as implied by the virial theorem. The magenta curve shows the zone of exclusion defined in
         \citet{Cappellari2013a}.
		}

\label{fig:sp-on-plane}
\end{figure*}

\begin{figure*}
\includegraphics[width=\textwidth]{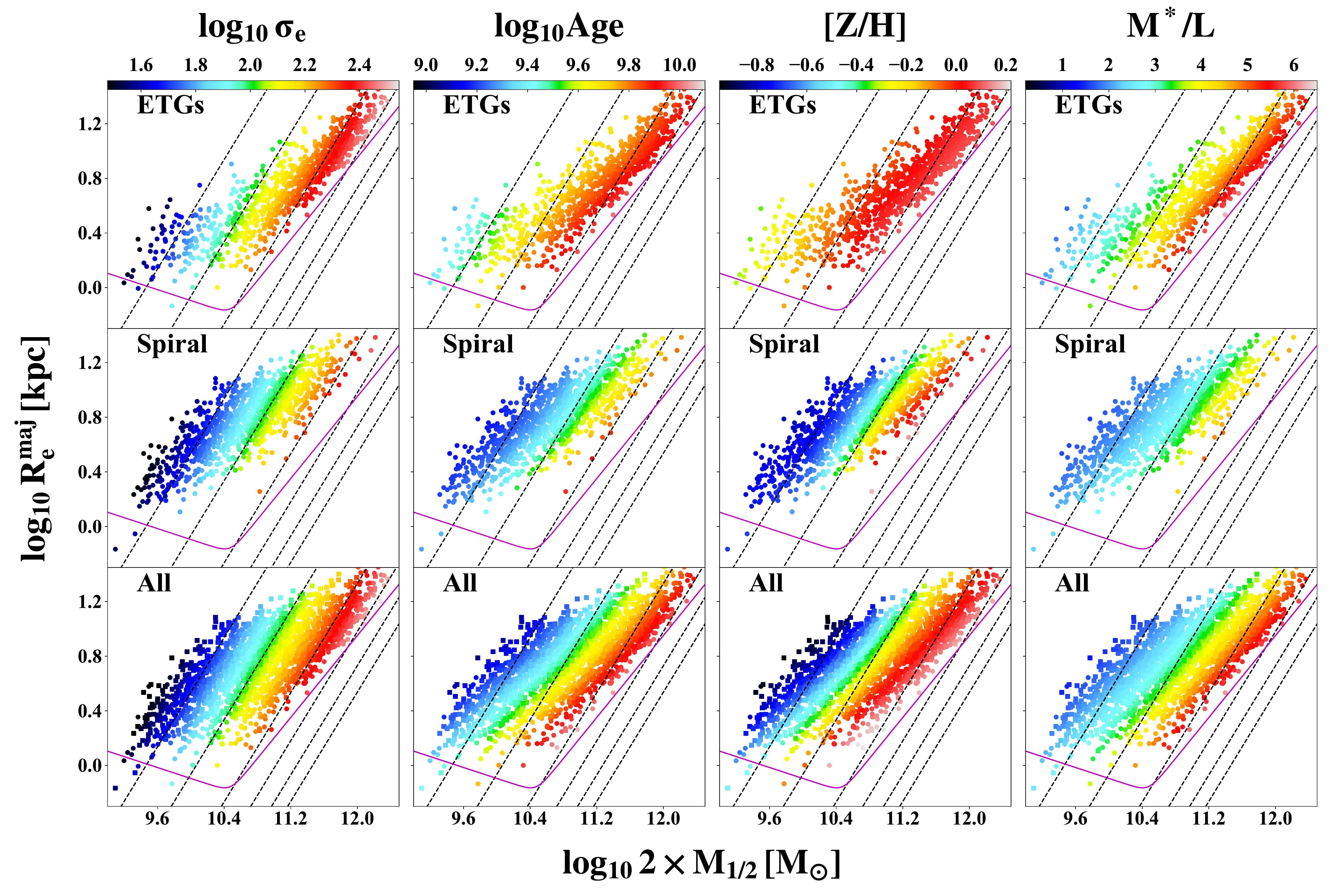}
\caption{The same as Fig~\ref{fig:sp-on-plane}, but with LOESS-smoothed $\sigma_{\rm e}$, logAge, [Z/H] and $M^*/L$.
		}

\label{fig:sp-on-plane_LOESS}
\end{figure*}

\begin{figure*}
\includegraphics[width=\textwidth]{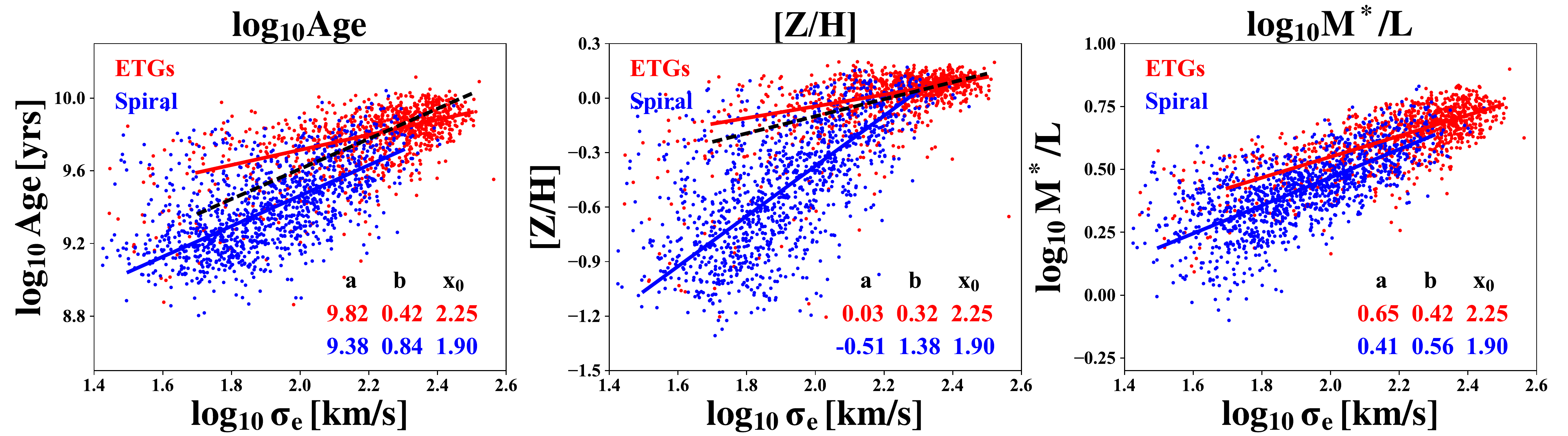}
\caption{logAge (left), metalicity (middle) and stellar mass-to-light ratio (right) vs. velocity dispersion. 
		 Early-type and spiral galaxies are shown with red and blue dots, respectively. The coloured solid lines
         show the best-fitting line $a+b(x-x_0)$ using the {\bf LTS\_LINEFIT} \citep{Cappellari2013b} procedure.
         The coefficients are shown in the lower right in each panel while the black dashed line shows the
         results from \citet{Scott2017} for their galaxies in clusters.}
\label{fig:1d-sigma}
\end{figure*}

\subsection{Stellar population gradient on the mass-size plane}
\label{sec:grident_on-plane}
We use the method described in Section~\ref{sec:sps} to estimate the age, 
metallicity and stellar mass-to-light ratio gradients for the galaxies in our sample. Similar
to Figs.~\ref{fig:sp-on-plane} and \ref{fig:sp-on-plane_LOESS}, we show the distribution of these
gradients on the mass-size plane in
Figs.~\ref{fig:gradient} and~\ref{fig:gradient_LOESS}. The systematic trends for the gradients 
are not as simple as the global properties which vary monotonically with the velocity dispersion 
as in Fig.~\ref{fig:sp-on-plane}, but there are several special
features for the distribution of the population gradients on the mass-size plane
\begin{enumerate}
 \item Many galaxies with small size and mass have positive age and stellar mass-to-light ratio gradient 
       ($\rm centre<outer$), which may be due to the star formation in galaxy centre
       \citep{Huang1996,Ellison2011,Oh2012},
       while nearly all the galaxies with $\log_{10} 2\times M_{1/2}>11.2$ have negative 
       age and stellar mass-to-light ratio gradients. 
       The age and the stellar mass-to-light ratio gradients for ETGs are correlated with galaxy mass.
 \item The metallicity gradients for spiral galaxies increase (become more negative) with mass and size,
       while for ETGs, the metallicity gradients change with velocity dispersions. 
       ETGs with higher and lower velocity dispersions have slightly shallower gradients.
 \item Spiral galaxies with large size and mass have the steepest age, metallicity and stellar mass-to-light
       ratio gradients. Although these galaxies are close to the massive ETGs on the mass-size plane, 
       their gradient properties have significant differences. One can see a clear boundary between these two 
       galaxy populations in Fig.~\ref{fig:gradient_LOESS}, especially for the metallicity gradient. 
       This may be due to the differences in their evolution histories, e.g. massive ETGs tend
       to have more mergers \citep{Cappellari2016}. This also agree with the scenario that many of them are slow 
       rotators (Graham et al. 2017, in preparation), which are thought to be formed by mergers 
       \citep{Naab2014,Penoyre2017,Li2018}.
\end{enumerate}
Here we note that our stellar mass-to-light gradients are based on the assumption of a constant
\citet{Salpeter1955} stellar initial mass function (IMF). The results will not be affected by the global
variation of the IMF (e.g. \citealt{Conroy2012}, \citealt{Cappellari2012}, \citealt{Li2017}), but by a
gradient of the IMF within galaxies (e.g. \citealt{vanDokkum2017} for massive elliptical galaxies).
The age and metallicity gradients will not be affected by the IMF.

\begin{figure*}
\includegraphics[width=0.9\textwidth]{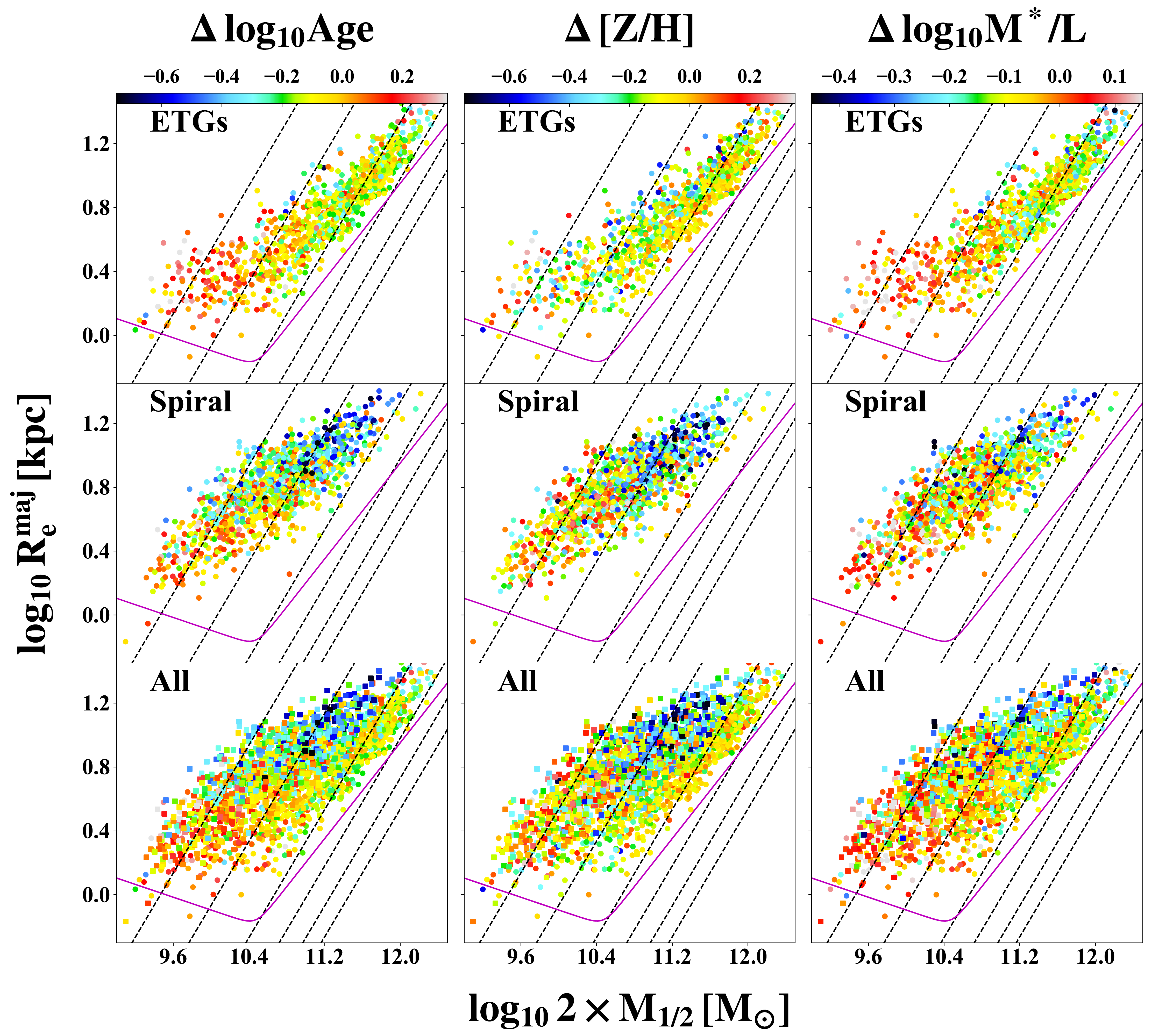}
\caption{Age, metallicity and stellar mass-to-light ratio gradient ($\Delta$logAge, $\Delta$[Z/H] and 
		 $\Delta M^*/L$) distribution on the mass-size plane. The
	     gradients are defined in Section~\ref{sec:sps}. A positive $\Delta$ value indicates a positive
         gradient, i.e. the central value is higher. Other labels are the same as in Fig~\ref{fig:sp-on-plane}.
         }
\label{fig:gradient}
\end{figure*}

\begin{figure*}
\includegraphics[width=0.9\textwidth]{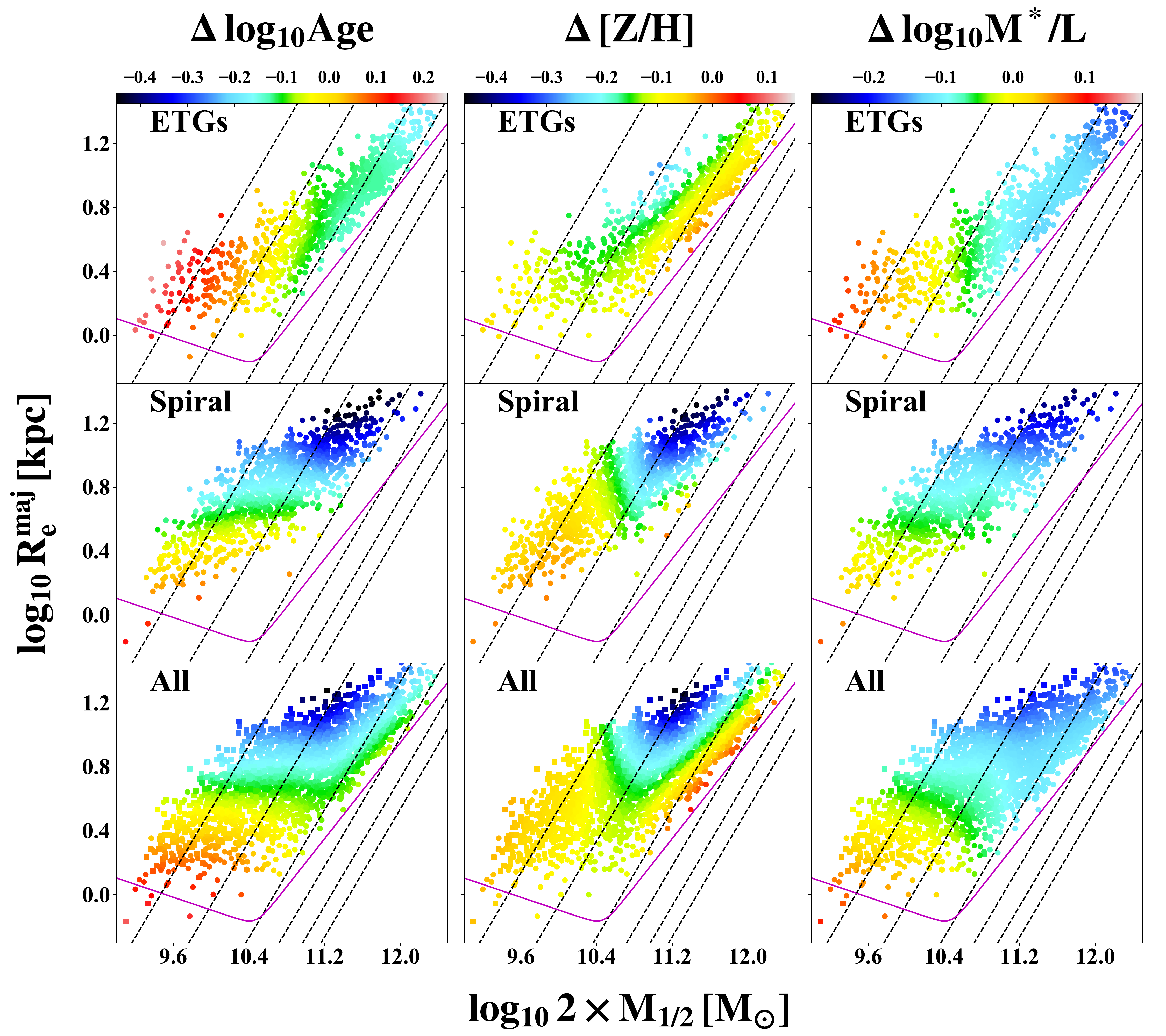}
\caption{The same as Fig~\ref{fig:gradient}, but with LOESS-smoothed $\Delta$logAge, $\Delta$[Z/H] and 
		 $\Delta M^*/L$.
		}
\label{fig:gradient_LOESS}
\end{figure*}

In Fig.~\ref{fig:1d-sigma-gradient}, we plot the histogram of the age, metallicity and
stellar mass-to-light ratio gradients and their relation with velocity dispersions.  
One can see that the age and the metallicity gradients peak around $\log \sigma_{e}=2.0$
(especially for metallicity gradients), which roughly corresponds to the critical mass
$\sim 3\times 10^{10}M_{\odot}$ of the break in the mass-size relation \citep{Cappellari2016},
below which no fully passive ETGs exist. The same critical mass is also shown in \citet{Kauffmann2003}.
This agree well with the results in \citet{Spolaor2009}, \citet{Tortora2010}, \citet{Kuntschner2010}
and \citet{Kuntschner2015}, but is slightly different from the results in \citet{Goddard2017},
which did not show a clear decrease of the metallicity gradients with increasing stellar masses.
The galaxies in our sample occupy similar region of the metallicity -- velocity dispersion relationship
with the simulated galaxies in \citet{Hopkins2009}. We also construct Figs~\ref{fig:gradient},
\ref{fig:gradient_LOESS} and \ref{fig:1d-sigma-gradient} with intermediate Sersic index (2-3) galaxies 
excluded, and find no significant differences from the main sample. 

In Fig.~\ref{fig:profile-on-plane}, to visually illustrate and confirm the reality of the the statistical
results of Figs.~\ref{fig:gradient} and \ref{fig:gradient_LOESS}, we show the metallicity profiles between 
$R_{\rm e}/8$ and $1R_{\rm e}$ of some galaxies with the best data qualities on the mass-size plane.
The selected galaxies have more than 400 Voronoi bins with S/N greater than 30 for the top 4 rows in
Fig.~\ref{fig:profile-on-plane}, and more than 200 Voronoi bins for the bottom row because small galaxies
have lower data qualities. The spiral galaxies with large size and mass have very steep metallicity gradients,
while the gradients for massive eliiptical galaxies are much shallower.
\begin{figure*}
\includegraphics[width=\textwidth]{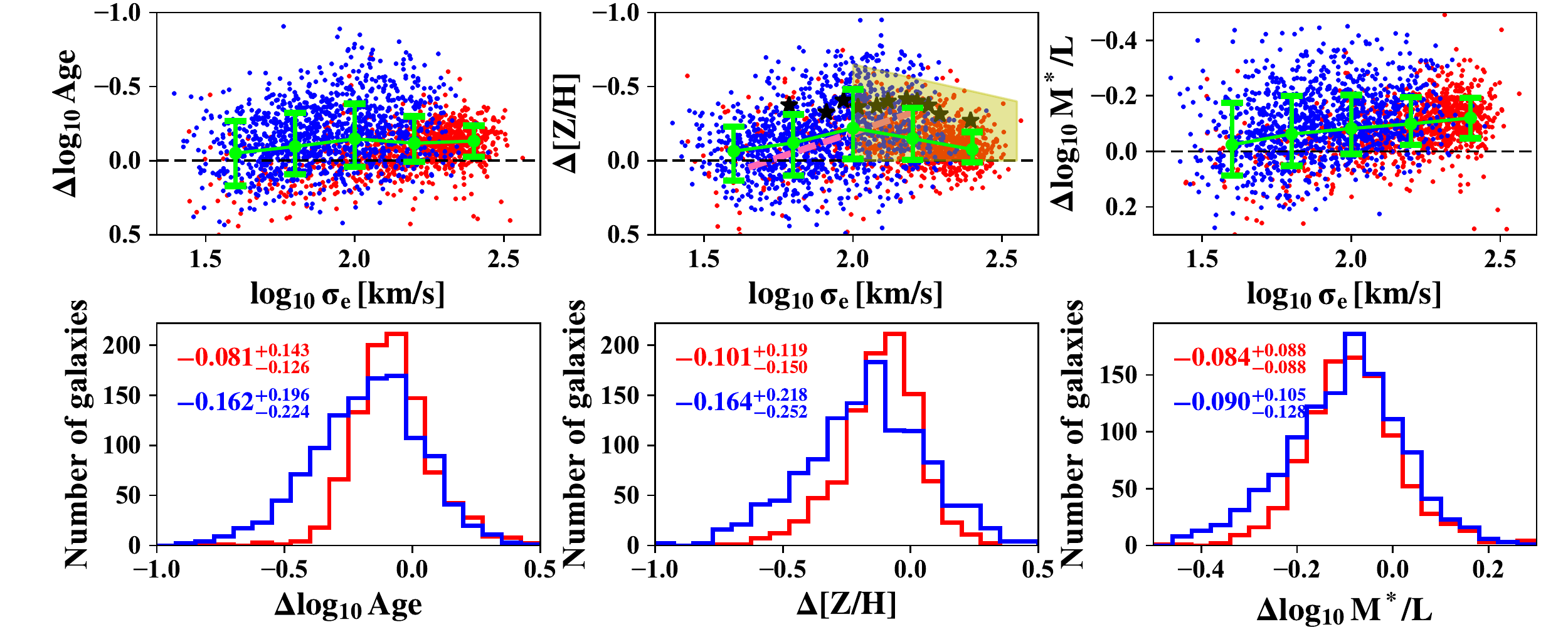}
\caption{Top: gradients of logAge (left), metalicity (middle) and stellar mass-to-light ratio (right)
         vs. velocity dispersion. Early-type and spiral galaxies are shown with red and blue dots, 
         respectively. The green error bars show the median and scatter of all the galaxies in each bin,
         calculated as the $16th$, $50th$ and $84th$ percentiles. In the middle panel, black stars are
         the results from \citet{Kuntschner2015}, the pink dashed line shows the trend from \citet{Spolaor2009}
         while the yellow shaded region shows the galaxy distribution from the simulation \citet{Hopkins2009}.
         Bottom: distribution of the logAge (left), metalicity (middle) and stellar mass-to-light ratio
         (right) gradients for the early-type and spiral galaxies.
         The medians and $1\sigma$ scatters of the distributions are shown in each panel.} 
\label{fig:1d-sigma-gradient}
\end{figure*}

\begin{figure*}
\includegraphics[width=0.8\textwidth]{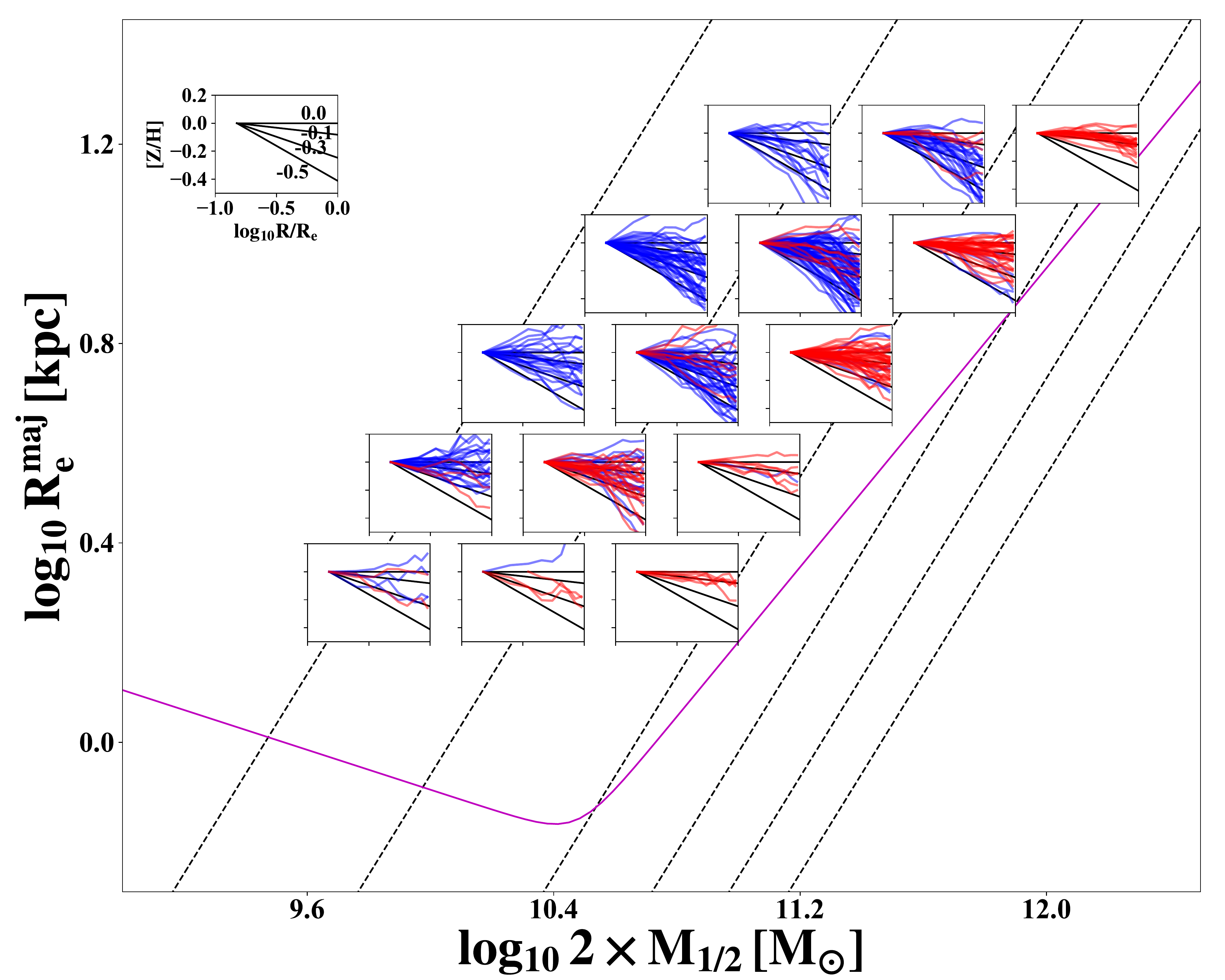}
\caption{Metallicity profiles ([Z/H] vs. $\log_{10}R/R_{\rm e}$) for the selected high S/N galaxies
         in different regions of the mass-size plane. Blue lines are the profiles for the spiral
         galaxies, red lines are for the ETGs. In each panel, the black solid lines
         represent the gradient values of $0$, $-0.1$, $-0.3$ and $-0.5$, respectively. All the panels
         have the same scales as shown in the upper left panel. All the profiles have been shifted to
         be 0 at $R_{\rm e}/8$.
		}
\label{fig:profile-on-plane}
\end{figure*}

\section{Conclusions}
\label{sec:conclusion}
Jeans Anisotropic modelling (JAM) and full spectrum fitting have been used to study the mass plane scaling 
relationship and the distribution of the stellar population properties on a inclined projection of the mass
plane, i.e. the mass-size plane. The galaxies used in this study are from the MaNGA DR14 sample, which is
currently the largest IFU sample including both early and late-type galaxies.

Below we summarize our main results:
\begin{enumerate}
 \item Both early-type and spiral galaxies are on a remarkably tight mass plane, with the best fitting
       coefficients close to the predicted values from the virial theorem. This extends the previous result
       for ETGs population to the whole galaxy population.
 \item The stellar population properties (i.e. age, metallicity and stellar mass-to-light ratio) of the galaxies
       in our sample vary systematically on the mass-size plane along roughly the velocity dispersion direction.
       The stellar population of the galaxies with higher velocity dispersion are older and more metal rich,
       which are consistent with a picture in which the bulge growth makes the population more metal rich
       and increases the likelihood for the star formation to be quenched \citep{Cappellari2016}.
 \item The gradients of age and stellar mass-to-light ratio could be positive ($\rm centre<outer$) for low 
       mass galaxies, while most massive galaxies have negative gradient.
 \item The metallicity-velocity dispersion relation shows a clear peck around 
       $\log \sigma_{\rm e}\approx 2.0$, which corresponds to the critical mass $\sim 3\times 10^{10}M_{\odot}$ of
       the break in the mass-size relation \citep{Cappellari2016}, below which no fully passive ETGs exist. 
 \item The distribution of the population gradients on the mass-size plane shows a clear boundary between massive
       spiral and early-type galaxies. Spiral galaxies with large size and mass have the steepest gradients. 
       In contrast, the massive ETGs located in similar region have shallower gradients.
       This may be due to differences in their evolution histories, e.g. mergers.
\end{enumerate}
The trends we see in this work, particularly the gradients shown in Figs. 6 and 7, are puzzling and warrant further studies. Observationally, these trends are still subject to the age-metallicity degeneracy and the scatters are still somewhat high. Theoretically, the spatial resolutions in numerical simulations and the treatments of physical processes, such as the chemical enrichment and feedback processes, will affect the final predictions of the stellar populations and their gradients. For example, according to \citet{Taylor2017} the metallicity gradients are affected by the initial steep gradients from gas-rich assembly, passive evolution by star formation and accretion at outskirts and flattening by mergers (major and minor). In reality, all these processes may be operating at the same time, and it will take further efforts to decode the information we assembled here. More detailed comparisons with high-resolution cosmological simulations such as Illustris \citep{Vogelsberger2014a,Vogelsberger2014b} and EAGLE \citep{Schaller2015} may be a fruitful next step.

\section*{Acknowledgements}
We thank Harald Kuntschner for providing the stellar population profiles for the 7 selected galaxies
from the $\rm ATLAS^{3D}$ survey and the referee for helpful comments. MC acknowledges support from a Royal Society University Research Fellowship.
We performed our computer runs on the Zen high performance computer cluster of the National
Astronomical Observatories, Chinese Academy of Sciences (NAOC), and the Venus server
at Tsinghua University. This work was supported by
the National Science Foundation of China (Grant No. 11333003, 11390372 to SM). 
This research made use of Marvin, a core Python package and web framework for MaNGA data,
developed by Brian Cherinka, Jos{\'e} S{\'a}nchez-Gallego, and Brett Andrews. (MaNGA Collaboration, 2017).

Funding for the Sloan Digital Sky Survey IV has been provided by
the Alfred P. Sloan Foundation, the U.S. Department of Energy Office of
Science, and the Participating Institutions. SDSS-IV acknowledges
support and resources from the Center for High-Performance Computing at
the University of Utah. The SDSS web site is www.sdss.org.

SDSS-IV is managed by the Astrophysical Research Consortium for the 
Participating Institutions of the SDSS Collaboration including the 
Brazilian Participation Group, the Carnegie Institution for Science, 
Carnegie Mellon University, the Chilean Participation Group, the French Participation Group, 
Harvard-Smithsonian Center for Astrophysics, 
Instituto de Astrof\'isica de Canarias, The Johns Hopkins University, 
Kavli Institute for the Physics and Mathematics of the Universe (IPMU) / 
University of Tokyo, Lawrence Berkeley National Laboratory, 
Leibniz Institut f\"ur Astrophysik Potsdam (AIP),  
Max-Planck-Institut f\"ur Astronomie (MPIA Heidelberg), 
Max-Planck-Institut f\"ur Astrophysik (MPA Garching), 
Max-Planck-Institut f\"ur Extraterrestrische Physik (MPE), 
National Astronomical Observatories of China, New Mexico State University, 
New York University, University of Notre Dame, 
Observat\'ario Nacional / MCTI, The Ohio State University, 
Pennsylvania State University, Shanghai Astronomical Observatory, 
United Kingdom Participation Group,
Universidad Nacional Aut\'onoma de M\'exico, University of Arizona, 
University of Colorado Boulder, University of Oxford, University of Portsmouth, 
University of Utah, University of Virginia, University of Washington, University of Wisconsin, 
Vanderbilt University, and Yale University.




\bibliographystyle{mnras}
\bibliography{lhy} 

\begin{thebibliography}{}
\makeatletter
\relax
\def\mn@urlcharsother{\let\do\@makeother \do\$\do\&\do\#\do\^\do\_\do\%\do\~}
\def\mn@doi{\begingroup\mn@urlcharsother \@ifnextchar [ {\mn@doi@}
  {\mn@doi@[]}}
\def\mn@doi@[#1]#2{\def\@tempa{#1}\ifx\@tempa\@empty \href
  {http://dx.doi.org/#2} {doi:#2}\else \href {http://dx.doi.org/#2} {#1}\fi
  \endgroup}
\def\mn@eprint#1#2{\mn@eprint@#1:#2::\@nil}
\def\mn@eprint@arXiv#1{\href {http://arxiv.org/abs/#1} {{\tt arXiv:#1}}}
\def\mn@eprint@dblp#1{\href {http://dblp.uni-trier.de/rec/bibtex/#1.xml}
  {dblp:#1}}
\def\mn@eprint@#1:#2:#3:#4\@nil{\def\@tempa {#1}\def\@tempb {#2}\def\@tempc
  {#3}\ifx \@tempc \@empty \let \@tempc \@tempb \let \@tempb \@tempa \fi \ifx
  \@tempb \@empty \def\@tempb {arXiv}\fi \@ifundefined
  {mn@eprint@\@tempb}{\@tempb:\@tempc}{\expandafter \expandafter \csname
  mn@eprint@\@tempb\endcsname \expandafter{\@tempc}}}

\bibitem[\protect\citeauthoryear{{Abolfathi} et~al.,}{{Abolfathi}
  et~al.}{2017}]{DR14}
{Abolfathi} B.,  et~al., 2017, preprint, \href
  {http://adsabs.harvard.edu/abs/2017arXiv170709322A} {} (\mn@eprint {arXiv}
  {1707.09322})

\bibitem[\protect\citeauthoryear{{Auger}, {Treu}, {Bolton}, {Gavazzi},
  {Koopmans}, {Marshall}, {Moustakas}  \& {Burles}}{{Auger}
  et~al.}{2010}]{Auger2010}
{Auger} M.~W.,  {Treu} T.,  {Bolton} A.~S.,  {Gavazzi} R.,  {Koopmans}
  L.~V.~E.,  {Marshall} P.~J.,  {Moustakas} L.~A.,   {Burles} S.,  2010,
  \mn@doi [\apj] {10.1088/0004-637X/724/1/511}, \href
  {http://adsabs.harvard.edu/abs/2010ApJ...724..511A} {724, 511}

\bibitem[\protect\citeauthoryear{{Blanton}, {Kazin}, {Muna}, {Weaver}  \&
  {Price-Whelan}}{{Blanton} et~al.}{2011}]{Blanton2011}
{Blanton} M.~R.,  {Kazin} E.,  {Muna} D.,  {Weaver} B.~A.,   {Price-Whelan} A.,
   2011, \mn@doi [\aj] {10.1088/0004-6256/142/1/31}, \href
  {http://adsabs.harvard.edu/abs/2011AJ....142...31B} {142, 31}

\bibitem[\protect\citeauthoryear{{Bolton}, {Burles}, {Treu}, {Koopmans}  \&
  {Moustakas}}{{Bolton} et~al.}{2007}]{Bolton2007}
{Bolton} A.~S.,  {Burles} S.,  {Treu} T.,  {Koopmans} L.~V.~E.,   {Moustakas}
  L.~A.,  2007, \mn@doi [\apjl] {10.1086/521357}, \href
  {http://adsabs.harvard.edu/abs/2007ApJ...665L.105B} {665, L105}

\bibitem[\protect\citeauthoryear{{Bruzual} \& {Charlot}}{{Bruzual} \&
  {Charlot}}{2003}]{Bruzual2003}
{Bruzual} G.,  {Charlot} S.,  2003, \mn@doi [\mnras]
  {10.1046/j.1365-8711.2003.06897.x}, \href
  {http://adsabs.harvard.edu/abs/2003MNRAS.344.1000B} {344, 1000}

\bibitem[\protect\citeauthoryear{{Bryant} et~al.,}{{Bryant}
  et~al.}{2015}]{Bryant2015}
{Bryant} J.~J.,  et~al., 2015, \mn@doi [\mnras] {10.1093/mnras/stu2635}, \href
  {http://adsabs.harvard.edu/abs/2015MNRAS.447.2857B} {447, 2857}

\bibitem[\protect\citeauthoryear{{Bundy} et~al.,}{{Bundy}
  et~al.}{2015}]{Bundy2015}
{Bundy} K.,  et~al., 2015, \mn@doi [\apj] {10.1088/0004-637X/798/1/7}, \href
  {http://adsabs.harvard.edu/abs/2015ApJ...798....7B} {798, 7}

\bibitem[\protect\citeauthoryear{{Calzetti}, {Armus}, {Bohlin}, {Kinney},
  {Koornneef}  \& {Storchi-Bergmann}}{{Calzetti} et~al.}{2000}]{Calzetti2000}
{Calzetti} D.,  {Armus} L.,  {Bohlin} R.~C.,  {Kinney} A.~L.,  {Koornneef} J.,
   {Storchi-Bergmann} T.,  2000, \mn@doi [\apj] {10.1086/308692}, \href
  {http://adsabs.harvard.edu/abs/2000ApJ...533..682C} {533, 682}

\bibitem[\protect\citeauthoryear{{Cappellari}}{{Cappellari}}{2002}]{Cappellari2002}
{Cappellari} M.,  2002, \mn@doi [\mnras] {10.1046/j.1365-8711.2002.05412.x},
  \href {http://adsabs.harvard.edu/abs/2002MNRAS.333..400C} {333, 400}

\bibitem[\protect\citeauthoryear{{Cappellari}}{{Cappellari}}{2008}]{Cappellari2008}
{Cappellari} M.,  2008, \mn@doi [\mnras] {10.1111/j.1365-2966.2008.13754.x},
  \href {http://adsabs.harvard.edu/abs/2008MNRAS.390...71C} {390, 71}

\bibitem[\protect\citeauthoryear{{Cappellari}}{{Cappellari}}{2016}]{Cappellari2016}
{Cappellari} M.,  2016, \mn@doi [\araa] {10.1146/annurev-astro-082214-122432},
  \href {http://adsabs.harvard.edu/abs/2016ARA%26A..54..597C} {54, 597}

\bibitem[\protect\citeauthoryear{{Cappellari}}{{Cappellari}}{2017}]{Cappellari2017}
{Cappellari} M.,  2017, \mn@doi [\mnras] {10.1093/mnras/stw3020}, \href
  {http://adsabs.harvard.edu/abs/2017MNRAS.466..798C} {466, 798}

\bibitem[\protect\citeauthoryear{{Cappellari} \& {Copin}}{{Cappellari} \&
  {Copin}}{2003}]{Cappellari2003}
{Cappellari} M.,  {Copin} Y.,  2003, \mn@doi [\mnras]
  {10.1046/j.1365-8711.2003.06541.x}, \href
  {http://adsabs.harvard.edu/abs/2003MNRAS.342..345C} {342, 345}

\bibitem[\protect\citeauthoryear{{Cappellari} \& {Emsellem}}{{Cappellari} \&
  {Emsellem}}{2004}]{Cappellari2004}
{Cappellari} M.,  {Emsellem} E.,  2004, \mn@doi [\pasp] {10.1086/381875}, \href
  {http://adsabs.harvard.edu/abs/2004PASP..116..138C} {116, 138}

\bibitem[\protect\citeauthoryear{{Cappellari} et~al.,}{{Cappellari}
  et~al.}{2006}]{Cappellari2006}
{Cappellari} M.,  et~al., 2006, \mn@doi [\mnras]
  {10.1111/j.1365-2966.2005.09981.x}, \href
  {http://adsabs.harvard.edu/abs/2006MNRAS.366.1126C} {366, 1126}

\bibitem[\protect\citeauthoryear{{Cappellari} et~al.,}{{Cappellari}
  et~al.}{2011}]{Cappellari2011}
{Cappellari} M.,  et~al., 2011, \mn@doi [\mnras]
  {10.1111/j.1365-2966.2010.18174.x}, \href
  {http://adsabs.harvard.edu/abs/2011MNRAS.413..813C} {413, 813}

\bibitem[\protect\citeauthoryear{{Cappellari} et~al.,}{{Cappellari}
  et~al.}{2012}]{Cappellari2012}
{Cappellari} M.,  et~al., 2012, \mn@doi [\nat] {10.1038/nature10972}, \href
  {http://adsabs.harvard.edu/abs/2012Natur.484..485C} {484, 485}

\bibitem[\protect\citeauthoryear{{Cappellari} et~al.,}{{Cappellari}
  et~al.}{2013a}]{Cappellari2013b}
{Cappellari} M.,  et~al., 2013a, \mn@doi [\mnras] {10.1093/mnras/stt562}, \href
  {http://adsabs.harvard.edu/abs/2013MNRAS.432.1709C} {432, 1709}

\bibitem[\protect\citeauthoryear{{Cappellari} et~al.,}{{Cappellari}
  et~al.}{2013b}]{Cappellari2013a}
{Cappellari} M.,  et~al., 2013b, \mn@doi [\mnras] {10.1093/mnras/stt644}, \href
  {http://adsabs.harvard.edu/abs/2013MNRAS.432.1862C} {432, 1862}

\bibitem[\protect\citeauthoryear{Cleveland \& Devlin}{Cleveland \&
  Devlin}{1988}]{Cleveland1988}
Cleveland W.~S.,  Devlin S.~J.,  1988, Journal of the American statistical
  association, 83, 596

\bibitem[\protect\citeauthoryear{{Conroy} \& {van Dokkum}}{{Conroy} \& {van
  Dokkum}}{2012}]{Conroy2012}
{Conroy} C.,  {van Dokkum} P.~G.,  2012, \mn@doi [\apj]
  {10.1088/0004-637X/760/1/71}, \href
  {http://adsabs.harvard.edu/abs/2012ApJ...760...71C} {760, 71}

\bibitem[\protect\citeauthoryear{{Di Matteo}, {Pipino}, {Lehnert}, {Combes}  \&
  {Semelin}}{{Di Matteo} et~al.}{2009}]{Matteo2009}
{Di Matteo} P.,  {Pipino} A.,  {Lehnert} M.~D.,  {Combes} F.,   {Semelin} B.,
  2009, \mn@doi [\aap] {10.1051/0004-6361/200911715}, \href
  {http://adsabs.harvard.edu/abs/2009A%26A...499..427D} {499, 427}

\bibitem[\protect\citeauthoryear{{Djorgovski} \& {Davis}}{{Djorgovski} \&
  {Davis}}{1987}]{Djorgovski1987}
{Djorgovski} S.,  {Davis} M.,  1987, \mn@doi [\apj] {10.1086/164948}, \href
  {http://adsabs.harvard.edu/abs/1987ApJ...313...59D} {313, 59}

\bibitem[\protect\citeauthoryear{{Dressler}, {Lynden-Bell}, {Burstein},
  {Davies}, {Faber}, {Terlevich}  \& {Wegner}}{{Dressler}
  et~al.}{1987}]{Dressler1987}
{Dressler} A.,  {Lynden-Bell} D.,  {Burstein} D.,  {Davies} R.~L.,  {Faber}
  S.~M.,  {Terlevich} R.,   {Wegner} G.,  1987, \mn@doi [\apj]
  {10.1086/164947}, \href {http://adsabs.harvard.edu/abs/1987ApJ...313...42D}
  {313, 42}

\bibitem[\protect\citeauthoryear{{Drory} et~al.,}{{Drory}
  et~al.}{2015}]{Drory2015}
{Drory} N.,  et~al., 2015, \mn@doi [\aj] {10.1088/0004-6256/149/2/77}, \href
  {http://adsabs.harvard.edu/abs/2015AJ....149...77D} {149, 77}

\bibitem[\protect\citeauthoryear{{Ellison}, {Nair}, {Patton}, {Scudder},
  {Mendel}  \& {Simard}}{{Ellison} et~al.}{2011}]{Ellison2011}
{Ellison} S.~L.,  {Nair} P.,  {Patton} D.~R.,  {Scudder} J.~M.,  {Mendel}
  J.~T.,   {Simard} L.,  2011, \mn@doi [\mnras]
  {10.1111/j.1365-2966.2011.19195.x}, \href
  {http://adsabs.harvard.edu/abs/2011MNRAS.416.2182E} {416, 2182}

\bibitem[\protect\citeauthoryear{{Emsellem}, {Monnet}  \& {Bacon}}{{Emsellem}
  et~al.}{1994}]{Emsellem1994}
{Emsellem} E.,  {Monnet} G.,   {Bacon} R.,  1994, \aap, \href
  {http://adsabs.harvard.edu/abs/1994A%26A...285..723E} {285, 723}

\bibitem[\protect\citeauthoryear{{Faber}, {Dressler}, {Davies}, {Burstein}  \&
  {Lynden-Bell}}{{Faber} et~al.}{1987}]{Faber1987}
{Faber} S.~M.,  {Dressler} A.,  {Davies} R.~L.,  {Burstein} D.,   {Lynden-Bell}
  D.,  1987, in {Faber} S.~M.,  ed., Nearly Normal Galaxies. From the Planck
  Time to the Present. pp 175--183

\bibitem[\protect\citeauthoryear{{Falc{\'o}n-Barroso},
  {S{\'a}nchez-Bl{\'a}zquez}, {Vazdekis}, {Ricciardelli}, {Cardiel}, {Cenarro},
  {Gorgas}  \& {Peletier}}{{Falc{\'o}n-Barroso}
  et~al.}{2011}]{Falcn-Barroso2011}
{Falc{\'o}n-Barroso} J.,  {S{\'a}nchez-Bl{\'a}zquez} P.,  {Vazdekis} A.,
  {Ricciardelli} E.,  {Cardiel} N.,  {Cenarro} A.~J.,  {Gorgas} J.,
  {Peletier} R.~F.,  2011, \mn@doi [\aap] {10.1051/0004-6361/201116842}, \href
  {http://adsabs.harvard.edu/abs/2011A%26A...532A..95F} {532, A95}

\bibitem[\protect\citeauthoryear{{Foreman-Mackey}, {Hogg}, {Lang}  \&
  {Goodman}}{{Foreman-Mackey} et~al.}{2013}]{Foreman2013}
{Foreman-Mackey} D.,  {Hogg} D.~W.,  {Lang} D.,   {Goodman} J.,  2013, \mn@doi
  [\pasp] {10.1086/670067}, \href
  {http://adsabs.harvard.edu/abs/2013PASP..125..306F} {125, 306}

\bibitem[\protect\citeauthoryear{{Goddard} et~al.,}{{Goddard}
  et~al.}{2017}]{Goddard2017}
{Goddard} D.,  et~al., 2017, \mn@doi [\mnras] {10.1093/mnras/stw2719}, \href
  {http://adsabs.harvard.edu/abs/2017MNRAS.465..688G} {465, 688}

\bibitem[\protect\citeauthoryear{{Graves} \& {Faber}}{{Graves} \&
  {Faber}}{2010}]{Graves2010}
{Graves} G.~J.,  {Faber} S.~M.,  2010, \mn@doi [\apj]
  {10.1088/0004-637X/717/2/803}, \href
  {http://adsabs.harvard.edu/abs/2010ApJ...717..803G} {717, 803}

\bibitem[\protect\citeauthoryear{{Graves}, {Faber}  \& {Schiavon}}{{Graves}
  et~al.}{2009}]{Graves2009}
{Graves} G.~J.,  {Faber} S.~M.,   {Schiavon} R.~P.,  2009, \mn@doi [\apj]
  {10.1088/0004-637X/698/2/1590}, \href
  {http://adsabs.harvard.edu/abs/2009ApJ...698.1590G} {698, 1590}

\bibitem[\protect\citeauthoryear{{Hopkins}, {Cox}, {Dutta}, {Hernquist},
  {Kormendy}  \& {Lauer}}{{Hopkins} et~al.}{2009}]{Hopkins2009}
{Hopkins} P.~F.,  {Cox} T.~J.,  {Dutta} S.~N.,  {Hernquist} L.,  {Kormendy} J.,
    {Lauer} T.~R.,  2009, \mn@doi [\apjs] {10.1088/0067-0049/181/1/135}, \href
  {http://adsabs.harvard.edu/abs/2009ApJS..181..135H} {181, 135}

\bibitem[\protect\citeauthoryear{{Huang}, {Gu}, {Su}, {Hawarden}, {Liao}  \&
  {Wu}}{{Huang} et~al.}{1996}]{Huang1996}
{Huang} J.~H.,  {Gu} Q.~S.,  {Su} H.~J.,  {Hawarden} T.~G.,  {Liao} X.~H.,
  {Wu} G.~X.,  1996, \aap, \href
  {http://adsabs.harvard.edu/abs/1996A%26A...313...13H} {313, 13}

\bibitem[\protect\citeauthoryear{{Hyde} \& {Bernardi}}{{Hyde} \&
  {Bernardi}}{2009}]{Hyde2009}
{Hyde} J.~B.,  {Bernardi} M.,  2009, \mn@doi [\mnras]
  {10.1111/j.1365-2966.2009.14783.x}, \href
  {http://adsabs.harvard.edu/abs/2009MNRAS.396.1171H} {396, 1171}

\bibitem[\protect\citeauthoryear{{Kauffmann} et~al.,}{{Kauffmann}
  et~al.}{2003}]{Kauffmann2003}
{Kauffmann} G.,  et~al., 2003, \mn@doi [\mnras]
  {10.1046/j.1365-8711.2003.06292.x}, \href
  {http://adsabs.harvard.edu/abs/2003MNRAS.341...54K} {341, 54}

\bibitem[\protect\citeauthoryear{{Koleva}, {Prugniel}, {De Rijcke}, {Zeilinger}
   \& {Michielsen}}{{Koleva} et~al.}{2009}]{Koleva2009}
{Koleva} M.,  {Prugniel} P.,  {De Rijcke} S.,  {Zeilinger} W.~W.,
  {Michielsen} D.,  2009, \mn@doi [Astronomische Nachrichten]
  {10.1002/asna.200911272}, \href
  {http://adsabs.harvard.edu/abs/2009AN....330..960K} {330, 960}

\bibitem[\protect\citeauthoryear{{Kuntschner}}{{Kuntschner}}{2015}]{Kuntschner2015}
{Kuntschner} H.,  2015, in {Cappellari} M.,  {Courteau} S.,  eds,  IAU
  Symposium Vol. 311, Galaxy Masses as Constraints of Formation Models. pp
  53--56, \mn@doi{10.1017/S1743921315003385}

\bibitem[\protect\citeauthoryear{{Kuntschner} et~al.,}{{Kuntschner}
  et~al.}{2010}]{Kuntschner2010}
{Kuntschner} H.,  et~al., 2010, \mn@doi [\mnras]
  {10.1111/j.1365-2966.2010.17161.x}, \href
  {http://adsabs.harvard.edu/abs/2010MNRAS.408...97K} {408, 97}

\bibitem[\protect\citeauthoryear{{La Barbera}, {Ferreras}, {de Carvalho},
  {Bruzual}, {Charlot}, {Pasquali}  \& {Merlin}}{{La Barbera}
  et~al.}{2012}]{Barbera2012}
{La Barbera} F.,  {Ferreras} I.,  {de Carvalho} R.~R.,  {Bruzual} G.,
  {Charlot} S.,  {Pasquali} A.,   {Merlin} E.,  2012, \mn@doi [\mnras]
  {10.1111/j.1365-2966.2012.21848.x}, \href
  {http://adsabs.harvard.edu/abs/2012MNRAS.426.2300L} {426, 2300}

\bibitem[\protect\citeauthoryear{{Lablanche} et~al.,}{{Lablanche}
  et~al.}{2012}]{Lablanche2012}
{Lablanche} P.-Y.,  et~al., 2012, \mn@doi [\mnras]
  {10.1111/j.1365-2966.2012.21343.x}, \href
  {http://adsabs.harvard.edu/abs/2012MNRAS.424.1495L} {424, 1495}

\bibitem[\protect\citeauthoryear{{Law} et~al.,}{{Law} et~al.}{2015}]{Law2015}
{Law} D.~R.,  et~al., 2015, \mn@doi [\aj] {10.1088/0004-6256/150/1/19}, \href
  {http://adsabs.harvard.edu/abs/2015AJ....150...19L} {150, 19}

\bibitem[\protect\citeauthoryear{{Law} et~al.,}{{Law} et~al.}{2016}]{Law2016}
{Law} D.~R.,  et~al., 2016, \mn@doi [\aj] {10.3847/0004-6256/152/4/83}, \href
  {http://adsabs.harvard.edu/abs/2016AJ....152...83L} {152, 83}

\bibitem[\protect\citeauthoryear{{Li}, {Li}, {Mao}, {Xu}, {Long}  \&
  {Emsellem}}{{Li} et~al.}{2016}]{Li2016}
{Li} H.,  {Li} R.,  {Mao} S.,  {Xu} D.,  {Long} R.~J.,   {Emsellem} E.,  2016,
  \mn@doi [\mnras] {10.1093/mnras/stv2565}, \href
  {http://adsabs.harvard.edu/abs/2016MNRAS.455.3680L} {455, 3680}

\bibitem[\protect\citeauthoryear{{Li} et~al.,}{{Li} et~al.}{2017}]{Li2017}
{Li} H.,  et~al., 2017, \mn@doi [\apj] {10.3847/1538-4357/aa662a}, \href
  {http://adsabs.harvard.edu/abs/2017ApJ...838...77L} {838, 77}

\bibitem[\protect\citeauthoryear{{Li}, {Mao}, {Emsellem}, {Xu}, {Springel}  \&
  {Krajnovi{\'c}}}{{Li} et~al.}{2018}]{Li2018}
{Li} H.,  {Mao} S.,  {Emsellem} E.,  {Xu} D.,  {Springel} V.,   {Krajnovi{\'c}}
  D.,  2018, \mn@doi [\mnras] {10.1093/mnras/stx2374}, \href
  {http://adsabs.harvard.edu/abs/2018MNRAS.473.1489L} {473, 1489}

\bibitem[\protect\citeauthoryear{{Lintott} et~al.,}{{Lintott}
  et~al.}{2008}]{Lintott2008}
{Lintott} C.~J.,  et~al., 2008, \mn@doi [\mnras]
  {10.1111/j.1365-2966.2008.13689.x}, \href
  {http://adsabs.harvard.edu/abs/2008MNRAS.389.1179L} {389, 1179}

\bibitem[\protect\citeauthoryear{{Lintott} et~al.,}{{Lintott}
  et~al.}{2011}]{Lintott2011}
{Lintott} C.,  et~al., 2011, \mn@doi [\mnras]
  {10.1111/j.1365-2966.2010.17432.x}, \href
  {http://adsabs.harvard.edu/abs/2011MNRAS.410..166L} {410, 166}

\bibitem[\protect\citeauthoryear{{Ma}, {Greene}, {McConnell}, {Janish},
  {Blakeslee}, {Thomas}  \& {Murphy}}{{Ma} et~al.}{2014}]{Ma2014}
{Ma} C.-P.,  {Greene} J.~E.,  {McConnell} N.,  {Janish} R.,  {Blakeslee} J.~P.,
   {Thomas} J.,   {Murphy} J.~D.,  2014, \mn@doi [\apj]
  {10.1088/0004-637X/795/2/158}, \href
  {http://adsabs.harvard.edu/abs/2014ApJ...795..158M} {795, 158}

\bibitem[\protect\citeauthoryear{{MacArthur}, {Gonz{\'a}lez}  \&
  {Courteau}}{{MacArthur} et~al.}{2009}]{MacArthur2009}
{MacArthur} L.~A.,  {Gonz{\'a}lez} J.~J.,   {Courteau} S.,  2009, \mn@doi
  [\mnras] {10.1111/j.1365-2966.2009.14519.x}, \href
  {http://adsabs.harvard.edu/abs/2009MNRAS.395...28M} {395, 28}

\bibitem[\protect\citeauthoryear{{McDermid} et~al.,}{{McDermid}
  et~al.}{2015}]{McDermid2015}
{McDermid} R.~M.,  et~al., 2015, \mn@doi [\mnras] {10.1093/mnras/stv105}, \href
  {http://adsabs.harvard.edu/abs/2015MNRAS.448.3484M} {448, 3484}

\bibitem[\protect\citeauthoryear{{Mehlert}, {Thomas}, {Saglia}, {Bender}  \&
  {Wegner}}{{Mehlert} et~al.}{2003}]{Mehlert2003}
{Mehlert} D.,  {Thomas} D.,  {Saglia} R.~P.,  {Bender} R.,   {Wegner} G.,
  2003, \mn@doi [\aap] {10.1051/0004-6361:20030886}, \href
  {http://adsabs.harvard.edu/abs/2003A%26A...407..423M} {407, 423}

\bibitem[\protect\citeauthoryear{{Naab} et~al.,}{{Naab}
  et~al.}{2014}]{Naab2014}
{Naab} T.,  et~al., 2014, \mn@doi [\mnras] {10.1093/mnras/stt1919}, \href
  {http://adsabs.harvard.edu/abs/2014MNRAS.444.3357N} {444, 3357}

\bibitem[\protect\citeauthoryear{{Navarro}, {Frenk}  \& {White}}{{Navarro}
  et~al.}{1996}]{Navarro1996}
{Navarro} J.~F.,  {Frenk} C.~S.,   {White} S.~D.~M.,  1996, \mn@doi [\apj]
  {10.1086/177173}, \href {http://adsabs.harvard.edu/abs/1996ApJ...462..563N}
  {462, 563}

\bibitem[\protect\citeauthoryear{{Oh}, {Oh}  \& {Yi}}{{Oh}
  et~al.}{2012}]{Oh2012}
{Oh} S.,  {Oh} K.,   {Yi} S.~K.,  2012, \mn@doi [\apjs]
  {10.1088/0067-0049/198/1/4}, \href
  {http://adsabs.harvard.edu/abs/2012ApJS..198....4O} {198, 4}

\bibitem[\protect\citeauthoryear{{Penoyre}, {Moster}, {Sijacki}  \&
  {Genel}}{{Penoyre} et~al.}{2017}]{Penoyre2017}
{Penoyre} Z.,  {Moster} B.~P.,  {Sijacki} D.,   {Genel} S.,  2017, \mn@doi
  [\mnras] {10.1093/mnras/stx762}, \href
  {http://adsabs.harvard.edu/abs/2017MNRAS.468.3883P} {468, 3883}

\bibitem[\protect\citeauthoryear{{Planck Collaboration} et~al.,}{{Planck
  Collaboration} et~al.}{2014}]{Planck2013}
{Planck Collaboration} et~al., 2014, \mn@doi [\aap]
  {10.1051/0004-6361/201321591}, \href
  {http://adsabs.harvard.edu/abs/2014A%26A...571A..16P} {571, A16}

\bibitem[\protect\citeauthoryear{{Rawle}, {Smith}  \& {Lucey}}{{Rawle}
  et~al.}{2010}]{Rawle2010}
{Rawle} T.~D.,  {Smith} R.~J.,   {Lucey} J.~R.,  2010, \mn@doi [\mnras]
  {10.1111/j.1365-2966.2009.15722.x}, \href
  {http://adsabs.harvard.edu/abs/2010MNRAS.401..852R} {401, 852}

\bibitem[\protect\citeauthoryear{{Roediger}, {Courteau}, {MacArthur}  \&
  {McDonald}}{{Roediger} et~al.}{2011}]{Roediger2011}
{Roediger} J.~C.,  {Courteau} S.,  {MacArthur} L.~A.,   {McDonald} M.,  2011,
  \mn@doi [\mnras] {10.1111/j.1365-2966.2011.19177.x}, \href
  {http://adsabs.harvard.edu/abs/2011MNRAS.416.1996R} {416, 1996}

\bibitem[\protect\citeauthoryear{{Roediger}, {Courteau},
  {S{\'a}nchez-Bl{\'a}zquez}  \& {McDonald}}{{Roediger}
  et~al.}{2012}]{Roediger2012}
{Roediger} J.~C.,  {Courteau} S.,  {S{\'a}nchez-Bl{\'a}zquez} P.,   {McDonald}
  M.,  2012, \mn@doi [\apj] {10.1088/0004-637X/758/1/41}, \href
  {http://adsabs.harvard.edu/abs/2012ApJ...758...41R} {758, 41}

\bibitem[\protect\citeauthoryear{Rousseeuw \& Van~Driessen}{Rousseeuw \&
  Van~Driessen}{2006}]{Rousseeuw2006}
Rousseeuw P.~J.,  Van~Driessen K.,  2006, Data mining and knowledge discovery,
  12, 29

\bibitem[\protect\citeauthoryear{{Salpeter}}{{Salpeter}}{1955}]{Salpeter1955}
{Salpeter} E.~E.,  1955, \mn@doi [\apj] {10.1086/145971}, \href
  {http://adsabs.harvard.edu/abs/1955ApJ...121..161S} {121, 161}

\bibitem[\protect\citeauthoryear{{S{\'a}nchez-Bl{\'a}zquez}
  et~al.,}{{S{\'a}nchez-Bl{\'a}zquez} et~al.}{2006a}]{Sanchez-Blazquez2006}
{S{\'a}nchez-Bl{\'a}zquez} P.,  et~al., 2006a, \mn@doi [\mnras]
  {10.1111/j.1365-2966.2006.10699.x}, \href
  {http://adsabs.harvard.edu/abs/2006MNRAS.371..703S} {371, 703}

\bibitem[\protect\citeauthoryear{{S{\'a}nchez-Bl{\'a}zquez}, {Gorgas}  \&
  {Cardiel}}{{S{\'a}nchez-Bl{\'a}zquez} et~al.}{2006b}]{Sanchez-Blazquez2006b}
{S{\'a}nchez-Bl{\'a}zquez} P.,  {Gorgas} J.,   {Cardiel} N.,  2006b, \mn@doi
  [\aap] {10.1051/0004-6361:20064846}, \href
  {http://adsabs.harvard.edu/abs/2006A%26A...457..823S} {457, 823}

\bibitem[\protect\citeauthoryear{{S{\'a}nchez-Bl{\'a}zquez}, {Forbes},
  {Strader}, {Brodie}  \& {Proctor}}{{S{\'a}nchez-Bl{\'a}zquez}
  et~al.}{2007}]{Sanchez-Blazquez2007}
{S{\'a}nchez-Bl{\'a}zquez} P.,  {Forbes} D.~A.,  {Strader} J.,  {Brodie} J.,
  {Proctor} R.,  2007, \mn@doi [\mnras] {10.1111/j.1365-2966.2007.11647.x},
  \href {http://adsabs.harvard.edu/abs/2007MNRAS.377..759S} {377, 759}

\bibitem[\protect\citeauthoryear{{S{\'a}nchez} et~al.,}{{S{\'a}nchez}
  et~al.}{2012}]{Sanchez2012}
{S{\'a}nchez} S.~F.,  et~al., 2012, \mn@doi [\aap]
  {10.1051/0004-6361/201117353}, \href
  {http://adsabs.harvard.edu/abs/2012A%26A...538A...8S} {538, A8}

\bibitem[\protect\citeauthoryear{{Schaller} et~al.,}{{Schaller}
  et~al.}{2015}]{Schaller2015}
{Schaller} M.,  et~al., 2015, \mn@doi [\mnras] {10.1093/mnras/stv1341}, \href
  {http://adsabs.harvard.edu/abs/2015MNRAS.452..343S} {452, 343}

\bibitem[\protect\citeauthoryear{{Schiavon}}{{Schiavon}}{2007}]{Schiavon2007}
{Schiavon} R.~P.,  2007, \mn@doi [\apjs] {10.1086/511753}, \href
  {http://adsabs.harvard.edu/abs/2007ApJS..171..146S} {171, 146}

\bibitem[\protect\citeauthoryear{{Scott} et~al.,}{{Scott}
  et~al.}{2015}]{Scott2015}
{Scott} N.,  et~al., 2015, \mn@doi [\mnras] {10.1093/mnras/stv1127}, \href
  {http://adsabs.harvard.edu/abs/2015MNRAS.451.2723S} {451, 2723}

\bibitem[\protect\citeauthoryear{{Scott} et~al.,}{{Scott}
  et~al.}{2017}]{Scott2017}
{Scott} N.,  et~al., 2017, preprint, \href
  {http://adsabs.harvard.edu/abs/2017arXiv170806849S} {} (\mn@eprint {arXiv}
  {1708.06849})

\bibitem[\protect\citeauthoryear{{S{\'e}rsic}}{{S{\'e}rsic}}{1963}]{Sersic1963}
{S{\'e}rsic} J.~L.,  1963, Boletin de la Asociacion Argentina de Astronomia La
  Plata Argentina, \href {http://adsabs.harvard.edu/abs/1963BAAA....6...41S}
  {6, 41}

\bibitem[\protect\citeauthoryear{{Smee} et~al.,}{{Smee}
  et~al.}{2013}]{Smee2013}
{Smee} S.~A.,  et~al., 2013, \mn@doi [\aj] {10.1088/0004-6256/146/2/32}, \href
  {http://cdsads.u-strasbg.fr/abs/2013AJ....146...32S} {146, 32}

\bibitem[\protect\citeauthoryear{{Spolaor}, {Proctor}, {Forbes}  \&
  {Couch}}{{Spolaor} et~al.}{2009}]{Spolaor2009}
{Spolaor} M.,  {Proctor} R.~N.,  {Forbes} D.~A.,   {Couch} W.~J.,  2009,
  \mn@doi [\apjl] {10.1088/0004-637X/691/2/L138}, \href
  {http://adsabs.harvard.edu/abs/2009ApJ...691L.138S} {691, L138}

\bibitem[\protect\citeauthoryear{{Springob} et~al.,}{{Springob}
  et~al.}{2012}]{Springob2012}
{Springob} C.~M.,  et~al., 2012, \mn@doi [\mnras]
  {10.1111/j.1365-2966.2011.19900.x}, \href
  {http://adsabs.harvard.edu/abs/2012MNRAS.420.2773S} {420, 2773}

\bibitem[\protect\citeauthoryear{{Taylor} \& {Kobayashi}}{{Taylor} \&
  {Kobayashi}}{2017}]{Taylor2017}
{Taylor} P.,  {Kobayashi} C.,  2017, \mn@doi [\mnras] {10.1093/mnras/stx1860},
  \href {http://adsabs.harvard.edu/abs/2017MNRAS.471.3856T} {471, 3856}

\bibitem[\protect\citeauthoryear{{Tortora} \& {Napolitano}}{{Tortora} \&
  {Napolitano}}{2012}]{Tortora2012}
{Tortora} C.,  {Napolitano} N.~R.,  2012, \mn@doi [\mnras]
  {10.1111/j.1365-2966.2012.20478.x}, \href
  {http://adsabs.harvard.edu/abs/2012MNRAS.421.2478T} {421, 2478}

\bibitem[\protect\citeauthoryear{{Tortora}, {Napolitano}, {Cardone},
  {Capaccioli}, {Jetzer}  \& {Molinaro}}{{Tortora} et~al.}{2010}]{Tortora2010}
{Tortora} C.,  {Napolitano} N.~R.,  {Cardone} V.~F.,  {Capaccioli} M.,
  {Jetzer} P.,   {Molinaro} R.,  2010, \mn@doi [\mnras]
  {10.1111/j.1365-2966.2010.16938.x}, \href
  {http://adsabs.harvard.edu/abs/2010MNRAS.407..144T} {407, 144}

\bibitem[\protect\citeauthoryear{{Vazdekis}, {S{\'a}nchez-Bl{\'a}zquez},
  {Falc{\'o}n-Barroso}, {Cenarro}, {Beasley}, {Cardiel}, {Gorgas}  \&
  {Peletier}}{{Vazdekis} et~al.}{2010}]{Vazdekis2010}
{Vazdekis} A.,  {S{\'a}nchez-Bl{\'a}zquez} P.,  {Falc{\'o}n-Barroso} J.,
  {Cenarro} A.~J.,  {Beasley} M.~A.,  {Cardiel} N.,  {Gorgas} J.,   {Peletier}
  R.~F.,  2010, \mn@doi [\mnras] {10.1111/j.1365-2966.2010.16407.x}, \href
  {http://adsabs.harvard.edu/abs/2010MNRAS.404.1639V} {404, 1639}

\bibitem[\protect\citeauthoryear{{Vogelsberger} et~al.,}{{Vogelsberger}
  et~al.}{2014a}]{Vogelsberger2014a}
{Vogelsberger} M.,  et~al., 2014a, \mn@doi [\mnras] {10.1093/mnras/stu1536},
  \href {http://adsabs.harvard.edu/abs/2014MNRAS.444.1518V} {444, 1518}

\bibitem[\protect\citeauthoryear{{Vogelsberger} et~al.,}{{Vogelsberger}
  et~al.}{2014b}]{Vogelsberger2014b}
{Vogelsberger} M.,  et~al., 2014b, \mn@doi [\nat] {10.1038/nature13316}, \href
  {http://adsabs.harvard.edu/abs/2014Natur.509..177V} {509, 177}

\bibitem[\protect\citeauthoryear{{Yan} et~al.,}{{Yan} et~al.}{2016a}]{Yan2016a}
{Yan} R.,  et~al., 2016a, \mn@doi [\aj] {10.3847/0004-6256/151/1/8}, \href
  {http://cdsads.u-strasbg.fr/abs/2016AJ....151....8Y} {151, 8}

\bibitem[\protect\citeauthoryear{{Yan} et~al.,}{{Yan} et~al.}{2016b}]{Yan2016b}
{Yan} R.,  et~al., 2016b, \mn@doi [\aj] {10.3847/0004-6256/152/6/197}, \href
  {http://cdsads.u-strasbg.fr/abs/2016AJ....152..197Y} {152, 197}

\bibitem[\protect\citeauthoryear{{York} et~al.,}{{York}
  et~al.}{2000}]{York2000}
{York} D.~G.,  et~al., 2000, \mn@doi [\aj] {10.1086/301513}, \href
  {http://adsabs.harvard.edu/abs/2000AJ....120.1579Y} {120, 1579}

\bibitem[\protect\citeauthoryear{{Zheng} et~al.,}{{Zheng}
  et~al.}{2015}]{Zheng2015}
{Zheng} Z.,  et~al., 2015, \mn@doi [\apj] {10.1088/0004-637X/800/2/120}, \href
  {http://adsabs.harvard.edu/abs/2015ApJ...800..120Z} {800, 120}

\bibitem[\protect\citeauthoryear{{Zheng} et~al.,}{{Zheng}
  et~al.}{2017}]{Zheng2017}
{Zheng} Z.,  et~al., 2017, \mn@doi [\mnras] {10.1093/mnras/stw3030}, \href
  {http://adsabs.harvard.edu/abs/2017MNRAS.465.4572Z} {465, 4572}

\bibitem[\protect\citeauthoryear{{van Dokkum}, {Conroy}, {Villaume}, {Brodie}
  \& {Romanowsky}}{{van Dokkum} et~al.}{2017}]{vanDokkum2017}
{van Dokkum} P.,  {Conroy} C.,  {Villaume} A.,  {Brodie} J.,   {Romanowsky}
  A.~J.,  2017, \mn@doi [\apj] {10.3847/1538-4357/aa7135}, \href
  {http://adsabs.harvard.edu/abs/2017ApJ...841...68V} {841, 68}

\makeatother
\end{thebibliography}



\appendix

\section{Example data table}

\clearpage
\begin{deluxetable}{lcccccccccc}
\tablewidth{0pt}
\tabletypesize{\small}
\setlength{\tabcolsep}{4pt}
\tablecaption{Properties of all the galaxies in the sample\label{tab:data}}
\tablehead{
 \colhead{MaNGA ID} &
 \colhead{Morphology} &
 \colhead{$\log_{10} \sigma_{\rm e}$} &
 \colhead{$\log_{10} M_{1/2}$} &
 \colhead{$\log_{10} R_{\rm e}^{\rm maj}$} &
 \colhead{$\log_{10}{\rm Age}$} &
 \colhead{[Z/H]} &
 \colhead{$\log_{10} M^*/L_{\rm r}$} &
 \colhead{$\Delta \log_{10} {\rm Age}$} &
 \colhead{$\Delta$[Z/H]} &
 \colhead{$\Delta \log_{10} M^*/L_{\rm r}$} \\
 \colhead{} &
 \colhead{} &
 \colhead{(km\,s$^{-1}$)} &
 \colhead{($M_{\odot}$)} &
 \colhead{(kpc)} &
 \colhead{(years)} &
 \colhead{} &
 \colhead{($M_{\odot}/L_{\odot \rm r}$)} &
 \colhead{} &
 \colhead{} &
 \colhead{} \\
 \colhead{(1)} &
 \colhead{(2)} &
 \colhead{(3)} &
 \colhead{(4)} &
 \colhead{(5)} &
 \colhead{(6)} &
 \colhead{(7)} &
 \colhead{(8)} &
 \colhead{(9)} &
 \colhead{(10)} &
 \colhead{(11)}
}
\startdata
1-320664     & S   &  2.17 & 10.66 &  0.76 &  9.95 &  0.06 &  0.75 & -0.15 &  0.14 & -0.08\\ 
1-321069     & E   &  2.37 & 11.48 &  1.07 &  9.80 &  0.14 &  0.66 & -0.12 & -0.05 & -0.12\\ 
1-235587     & E   &  2.04 & 10.27 &  0.50 &  9.66 &  0.06 &  0.54 &  0.04 & -0.33 & -0.04\\ 
1-320677     & E   &  2.27 & 11.41 &  1.14 &  9.70 &  0.12 &  0.60 & -0.38 &  0.18 & -0.30\\ 
1-235576     & S   &  2.10 & 10.82 &  0.75 &  9.74 & -0.57 &  0.68 & -0.04 &  0.45 & -0.14\\ 
1-235530     & E   &  2.22 & 10.53 &  0.41 &  9.90 &  0.07 &  0.73 & -0.33 & -0.15 & -0.28\\ 
1-235398     & S   &  2.10 & 10.67 &  0.78 &  9.71 & -0.21 &  0.57 &  0.13 & -0.10 &  0.06\\ 
1-320606     & S   &  1.86 & 10.46 &  0.93 &  9.04 & -0.86 &  0.21 & -0.43 & -0.55 & -0.31\\ 
1-321074     & E   &  2.27 & 11.04 &  0.83 &  9.64 &  0.15 &  0.56 & -0.10 & -0.09 & -0.10\\ 
1-235582     & E   &  1.84 &  9.92 &  0.58 &  9.33 &  0.07 &  0.30 &  0.31 & -0.20 &  0.26\\ 
1-320584     & E   &  2.47 & 11.70 &  1.08 &  9.94 &  0.12 &  0.77 & -0.19 & -0.06 & -0.16\\ 
1-235611     & S   &  2.06 & 11.00 &  1.11 &  9.56 & -0.15 &  0.56 & -0.68 & -0.52 & -0.39\\ 
1-320655     & E   &  2.37 & 11.25 &  0.78 &  9.85 &  0.12 &  0.70 & -0.23 &  0.00 & -0.16\\ 
1-24092      & E   &  2.13 & 10.37 &  0.35 &  9.01 & -0.73 &  0.42 &  0.08 & -0.28 &  0.35\\ 
1-23979      & E   &  2.00 & 10.25 &  0.53 &  9.59 & -0.10 &  0.53 &  0.02 & -0.22 & -0.01\\ 
1-24099      & E   &  2.07 & 10.23 &  0.38 &  9.68 &  0.10 &  0.57 & -0.11 & -0.17 & -0.08\\ 
1-23929      & S   &  1.80 & 10.25 &  0.83 &  9.34 & -0.52 &  0.41 & -0.41 & -0.12 & -0.08\\ 
1-24368      & S   &  1.72 &  9.96 &  0.77 &  9.19 & -0.76 &  0.20 & -0.14 &  0.06 & -0.33\\ 
1-24354      & S   &  1.75 &  9.87 &  0.51 &  9.83 & -0.26 &  0.56 & -0.14 &  0.10 & -0.07\\ 
1-595027     & S   &  2.05 & 10.72 &  0.90 &  9.24 & -0.75 &  0.48 & -0.25 & -0.72 & -0.04\\ 
1-595093     & S   &  2.17 & 10.99 &  1.03 &  9.66 & -0.02 &  0.64 & -0.40 & -0.13 & -0.19\\ 
1-24018      & E   &  2.11 & 10.58 &  0.71 &  9.78 &  0.07 &  0.64 & -0.11 & -0.22 & -0.14\\ 
1-23891      & S   &  2.20 & 10.82 &  0.78 &  9.80 &  0.03 &  0.66 & -0.26 &  0.06 & -0.18\\ 
1-24148      & S   &  2.10 & 10.61 &  0.71 &  9.81 &  0.05 &  0.69 & -0.19 &  0.10 & -0.13\\ 
1-25937      & S   &  2.22 & 11.34 &  1.22 &  9.70 & -0.00 &  0.60 & -0.36 & -0.39 & -0.25\\ 
1-25911      & E   &  2.37 & 11.57 &  1.14 &  9.72 &  0.09 &  0.59 & -0.24 & -0.09 & -0.23\\ 
1-24124      & S   &  1.68 &  9.74 &  0.53 &  9.43 &  0.04 &  0.43 &  0.06 &  0.07 & -0.02\\ 
1-115062     & E   &  2.12 & 10.02 &  0.00 &  9.80 &  0.11 &  0.65 & -0.00 &  0.04 & -0.02\\ 
1-114928     & E   &  2.24 & 10.76 &  0.67 &  9.91 & -0.02 &  0.71 & -0.12 & -0.31 & -0.19\\ 
1-115128     & S   &  1.99 & 10.70 &  0.94 &  9.23 & -0.59 &  0.34 & -0.16 & -0.50 & -0.15\\ 
\enddata
\tablecomments{
Column (1): The MaNGA ID of the galaxy. 
Column (2): Galaxy morphology. E for ETGs, S for spiral galaxies.
Column (3): Velocity dispersion within $1R_{\rm e}$, as defined in equation~\ref{eq:sigma}.
Column (4): Enclosed total mass within 3-dimensional half-light radius from dynamical model, $M_{1/2}$.
Column (5): Major axis of the half-light isophote for the best fitting MGE model.
Column (6): Mean logAge within the effective radius.
Column (7): Mean metaliticy within the effective radius.
Column (8): Mean stellar mass-to-light ratio within the effective radius in SDSS r-band.
Column (9): Age gradient.
Column (10): Metalicity gradient.
Column (11): Stellar mass-to-light ratio gradient in SDSS r-band.
Please see the journal website for the complete table.
}
\end{deluxetable}

\bsp	
\label{lastpage}
\end{document}